\documentclass[12pt, pra, groupedaddress, amsmath,amssymb, nofootinbib, superscriptaddress]{revtex4-2}
\usepackage{graphicx}
\usepackage{dcolumn}
\usepackage{mathrsfs}
\usepackage{amssymb}
\usepackage{amsthm}
\usepackage{svg}
\usepackage{bm}
\usepackage{hyperref}
\usepackage{subcaption}
\usepackage{mathtools}
\usepackage{multirow}
\hypersetup{
 colorlinks=true,
 citecolor=blue,
 linkcolor=red,
 urlcolor=blue,
 pdfpagemode=UseNone,
 pdfstartview=FitH}
\usepackage{lipsum, physics}

\begin{document}
\newcommand{\omS}[1]{\omega_{\mathcal{S}_{#1}}}
\newcommand{\omSS}{\omega_{\mathcal{S}_{1}\mathcal{S}_{2}}}
\newcommand{\omE}[1]{\omega_{\mathcal{E}_{#1}}}
\newcommand{\omES}[2]{\omega_{\mathcal{E}_{#1}\mathcal{S}_{#2}}}
\newcommand{\sigS}[1]{\hat{\sigma}^{z}_{\mathcal{S}_{#1}}}
\newcommand{\sigE}[1]{\hat{\sigma}^{z}_{\mathcal{E}_{#1}}}
\newcommand{\sigES}[2]{\hat{\sigma}^{z}_{\mathcal{E}_{#1}\mathcal{S}_{#2}}}
\newcommand{\g}[1]{\Gamma_{\mathcal{E}_{k}\mathcal{S}_{#1}}}
\newcommand{\gS}[1]{\Gamma_{\mathcal{S}_{#1}}}
\newcommand{\gES}[2]{\Gamma_{\mathcal{E}_{#1}\mathcal{S}_{#2}}}
\newcommand{\gl}[1]{\Gamma_{\mathcal{S}_{#1}\mathcal{E}_{l}}}
\newcommand{\gk}[1]{\Gamma_{\mathcal{S}_{#1}\mathcal{E}_{k}}}
\newcommand{\gHomo}[1]{\Gamma_{\mathcal{S}_{#1}}}
\newcommand{\E}{\mathcal{E}}
\newcommand{\sumk}{\sum_{k=1}^{N}}
\newcommand{\s}[1]{s_{#1}(t)}
\newcommand{\scon}[1]{s^{*}_{#1}(t)}
\newcommand{\bigO}{\bigotimes_{k=1}^{N}}
\newcommand{\En}[1]{\psi_{\mathcal{E}{#1}}(t)}
\newcommand{\rS}[2]{r_{\mathcal{S}_{#1}#2}(t)}
\newcommand{\rScon}[2]{r^{*}_{\mathcal{S}_{#1}#2}(t)}
\newcommand{\rSS}[1]{r_{\mathcal{S}_{1}\mathcal{S}_{2}#1}(t)}
\newcommand{\rSScon}[1]{r^{*}_{\mathcal{S}_{1}\mathcal{S}_{2}#1}(t)}
\newcommand{\expo}[1]{\text{exp}\left(#1\right)}
\newcommand{\expb}[1]{\text{exp}[#1]}
\newcommand{\pp}[1]{\sum_{p_{1}=0}^{1}\cdots\sum_{p_{N}=0}^{1}\prescript{}{\E}{\bra{p_{1},\cdots, p_{N}}}{#1} \ket{p_{1},\cdots, p_{N}}_{\E}}
\newcommand{\pps}[1]{\sum_{p_{k}=0}^{1}\prescript{}{k}{\bra{p_{k}}}{#1}\ket{p_{k}}_{k}}
\raggedbottom

\title{Unique entanglement time evolution of two-qubit product separable and extended Werner-like states in a discrete qubit environment}

\author{Renzo P. Alporha}
\affiliation{Department of Physics, Mapúa University, Intramuros, Manila, Philippines}
\author{Lemuel John F. Sese}
\affiliation{Department of Physics, Pohang University of Science and Technology, Pohang, Korea}
\author{Rayda P. Gammag}
\email{rpgammag@mapua.edu.ph}
\affiliation{Department of Physics, Mapúa University, Intramuros, Manila, Philippines}

\begin{abstract}
\textcolor{black}{This study investigates the parameters affecting the entanglement time evolution of product separable (PS) and extended Werner-like (EWL) states in homogeneous, white noise, and mixed environments. In a pure homogeneous environment, both states demonstrate complete entanglement revivals, where an increase in the number of environments leads to an attenuation of concurrence. The PS state exhibits gaps and never reaches a maximum entanglement, whereas maximum purity EWL states (Bell states) maintain or periodically reach maximum entanglement. Hence, the PS and EWL states have a disjoint entanglement time evolution subspaces. Interestingly, the environment interaction and subsystem coupling interaction that influence entanglement have an inverse time relationship under a constant value of concurrence. Placing the system of interest in a white noise environment induces entanglement dissipation, with the dissipation time dependent on the width of random distribution of interaction strengths rather than the magnitude of interaction strength. In a distinct qubit environment, the entanglement time evolution of the PS state depends on the distribution of the number of environments. Moreover, combining homogeneous and white noise environments results in entanglement dynamics that exhibit characteristics of both homogeneous and white noise.}
\end{abstract}


\maketitle
\section{Introduction}
\textcolor{black}{Entanglement is a quantum phenomenon that arises when the interaction between two particles transforms the two-particle wavefunction into one inseparable wavefunction that can no longer be described by its local constituents \cite{Schrodinger_1935}. This strange property, combined with the probabilistic nature of quantum systems, finds application in quantum cryptography \cite{Ekert1991, Naik2000, Tittel2000}, quantum dense coding \cite{Mattle1996, Fang2000, Mizuno2005, Jing2003}, quantum teleportation \cite{Bennett1993, Bouwmeester1997, Sherson2006}, and quantum algorithms \cite{Linden2002, Orus2004, Horodecki_2009}.} 

\textcolor{black}{Entanglement is not the reason for the increase in the computation speed of a quantum computer as articulated by the Gottesman–Knill theorem \cite{gottesman1998, Nielsen_et_al_2010}. This theorem shows that many quantum algorithms relying on entanglement can be efficiently simulated classically as well \cite{Linden2002}.}

\textcolor{black}{One of the simplest systems that exhibits entanglement is a two-qubit system. For it to be entangled, the two qubits must initially interact with each other. But having a two-qubit system interact with an environment poses significant challenges to its practical realization and applications \cite{Zyczkowski_2001}. For instance, coupling a pure bipartite system to a thermal reservoir invariably leads to entanglement decay \cite{Tahira2008}. A study shows quasi-periodic revival of entanglement for an arbitrary state in noisy environments but it still leads to asymptotic decay \cite{Shi2016}. In addition, some initially entangled states subjected to vacuum noise exhibit asymptotic entanglement decay, indicating a partial sudden death \cite{Yu2004, Tahira2008}. However, there are some entangled states that can be preserved with the appropriate initial condition and qubit systems configuration even in the presence of an environment \cite{Flores2015}. The type of environment that is considered in these studies is the usual harmonic oscillator bath.} 

\textcolor{black}{One may ask about the possible effects of two-qubit entanglement when the system interacts with a qubit bath. Although not related to entanglement, the model used by Cucchietti and Zurek is worth noting. \cite{Cucchietti2005}. Their work determines a Gaussian decay of coherences of a single qubit (spin) interacting with a qubit environment. A similar study by Sese and Galapon \cite{Sese2022} investigated a qubit that interacts with a discrete homogeneous environment and demonstrated that the system's coherence can periodically recover its dissipated information, resulting in an exact recurrence. Given these results, we investigate the possible revival of entanglement in a two-qubit system interacting with a qubit bath.} 

\textcolor{black}{Gedik \cite{Gedik2006} employed an interaction Hamiltonian similar to that of Cucchietti and Zurek. He used it to study the effect of a spin bath in Bell states. Gedik obtained a similar result: the coherence of a Bell state has a Gaussian decay.  However, the results of Gedik focused more on the violation of Bell inequality in coherent and decoherent regimes. What distinguishes this current work from the previously mentioned articles is the emphasis on the parameters of the environment that affect the time evolution of the entangled two-qubit system.}

\textcolor{black}{With the previous studies presented, we are motivated to investigate the entanglement time evolution of a two-qubit immersed in a $N$-qubit environment. In our preliminary work \cite{Alporha_2024}, we showed that it is possible for a periodic entanglement revival to occur provided that the environment interaction is rationally dependent on the system's internal frequencies. This is the same condition presented by Galapon and Sese to obtain the revival of coherences \cite{Sese2022}. However, our preliminary studies only involved a product-separable (PS) initial state for the system and environment. This led us to extend our prior study to also investigate the entanglement time evolution of a two-qubit system initially prepared in extended Werner-like (EWL) states. Having an initially prepared EWL state is worth exploring since it manifests both Werner states (mixed states) and Bell-like states (pure, entangled states). This work aims to identify the influence of the parameters in the time evolution of the entanglement of our model, providing relevant insights into the manipulation and preservation of the entanglement. Concurrence will be used to show how entanglement will behave (preserve or decay) interacting with $N$-qubit environment with different initial states. The states will be examined in three different types of environments: homogeneous, white noise, and mixed. In a homogeneous environment, all qubits interact with the same coupling strength. In contrast, a white noise environment assumes that every qubit interaction strength is randomly selected from a uniform distribution. The mixed environment combines these two scenarios. We then derive the reduced density matrix $\rho_{\mathcal{S}}$ as a function of time $t$ to calculate the concurrence, which in turn determines the entanglement time evolution of the system.}

\textcolor{black}{The results reveal that the entanglement time evolution of the states is highly contingent on the initial preparation. Specifically, a homogeneous environment induces revivals and attenuation in the entanglement dynamics, whereas a white noise environment causes entanglement decay. The mixed environment combines the characteristics of both homogeneous and white noise environments. We also show the distinction of the entanglement time evolution for different initial states (PS and EWL) with the same setup of environment.} 

\textcolor{black}{The paper is arranged as follows: we introduce the two initial states, specifically PS and EWL states in Sec. \ref{Initial state preparation}. We then demonstrate their entanglement time evolution for both mutual and distinct qubit environments in Secs. \ref{Mutual qubit environment} and \ref{Distinct Environment}, respectively. Lastly, a conclusion is provided in Sec. \ref{Conclusion}.}

\section{Initial states}\label{Initial state preparation}
\textcolor{black}{For this study, the time-evolved state of the system of interest (SOI) with a Hilbert space $\mathcal{H_S}=\mathbb{C}_{\mathcal{S}_1}^2\otimes\mathbb{C}_{\mathcal{S}_2}^2$ can be represented by the time-evolved reduced density matrix, 
\begin{equation}
\label{rho_S(t)}
    \rho_{\mathcal{S}}(t)=\sum_{l,m,n,o=0}^{1}a_{lmno}r_{lmno}(t)\ket{lm}\bra{no},
\end{equation}
\noindent where $a_{lmno}$'s are coefficients that is obtained from the initial conditions. Moreover, $r_{lmno}(t)$ is the time-evolved parameter that dictates the time-evolution of the system. This parameter will be dependent on the initial state and the parameters (e.g. two-qubit interaction coupling, interaction coupling with the environment, and self-dynamical parameter) of the Hamiltonian that is to be considered. The off-diagonal elements ($l\neq n$ and $m\neq o$) are considered as the interference of the mutually exclusive bases of SOI. These are commonly known as coherences. Since we consider that SOI interacts with an environment (to be discussed in Secs.~\ref{Mutual qubit environment} and \ref{Distinct Environment}), this time-evolved parameter decays, hence it is called decoherence factor. Usually decoherence is considered as the dissipation of information from the system to the environment \cite{Zurek1991}.} 

\textcolor{black}{We consider that SOI was initially prepared in either PS state or EWL state. These two states represent the boundaries of entanglement. Here, PS state indicates an absence of entanglement, while EWL can represent maximum entanglement. Concurrence is used to quantify the entanglement of the SOI dictated by the time-evolved state in Eq.~(\ref{rho_S(t)}). Concurrence is a standard quantitative measurement of entanglement for two qubits and is derived from the entanglement of formation as a measure criterion. In other words, it also measures how far the states are from being separable. The generalized concurrence formula is given by Hill and Wootters \cite{Hill_1997,Wootters_1998},
\begin{equation}
\label{C_formula}
    C(\rho_{\mathcal{S}})=\text{max}\{0,\nu_{1}-\nu_{2}-\nu_{3}-\nu_{4}\},
\end{equation}
where $\nu_{j}$ for all $j$ denote the square roots of the eigenvalues of $\rho_{\mathcal{S}}\tilde{\rho_{\mathcal{S}}}$ in decreasing order, with $\nu_{1}$ being the highest.  Here, $\tilde{\rho}_{\mathcal{S}}$ is the spin-flipped state. To calculate $\tilde{\rho_{\mathcal{S}}}$, the spin-flip operation is applied to the reduced density matrix $\rho_{\mathcal{S}}$, which is defined as
\begin{equation}
    \tilde{\rho_\mathcal{S}}=(\sigma^{y}\otimes\sigma^{y})\rho^{*}(\sigma^{y}\otimes\sigma^{y}).
\end{equation}
Since the parameter $r_{lmno}(t)$ dictates the time evolution of $\rho_{\mathcal{S}}$, the time evolution of concurrence is only dependent on this parameter. The explicit form of $r_{lmno}(t)$ will be the input to calculate the concurrence numerically using Matlab.}

\subsection{Product separable state}\label{Product separable state}
\textcolor{black}{A general pure state $\ket{\Psi}\in\mathcal{H_S}$ can be expressed as, 
\begin{equation}
\label{Sep_Ini}
    \ket{\Psi}=a_{00}\ket{00}+a_{01}\ket{01}+a_{10}\ket{10}+a_{11}\ket{11}.
\end{equation}
Considering two subsystems $\mathbb{C}_{\mathcal{S}_1}^2$ and $\mathbb{C}_{\mathcal{S}_2}^2$ that do not interact initially ($t=0$), this means that both subsystems are closed quantum systems; hence, their states are pure and separable (the state of SOI can be described by the states of each subsystem independently). For a bipartite system, we define a product-separable (PS) state (equivalent to being unentangled) if it can be expressed as a convex combination of tensor products of two pure states from sub-Hilbert space orthonormal bases, that is,
\begin{equation}\label{PSS_1}
    \ket{\Psi(0)}_{PS}=\left(a_{0}^{(1)}\ket{0}_{1}+a_{1}^{(1)}\ket{1}_{1}\right) \otimes \left(a_{0}^{(2)}\ket{0}_{2}+a_{1}^{(2)}\ket{1}_{2}\right).
\end{equation}
Here $\ket{i}_j$ represents the orthonormal basis $i$ in $j$ subsystem. The probability amplitudes will be considered unbiased, i.e. $a_{i}^{(j)}=1/\sqrt{2}$ for all $i$ and $j$. The general pure state in Eq.~(\ref{Sep_Ini}) will be equivalent to Eq.~(\ref{PSS_1}) if $a_{lm}=a_{l}^{(1)}a_{m}^{(2)}$.}

\subsection{Extended Werner-like states}\label{Extended Werner-like states}
\textcolor{black}{Another initial state that can be considered is the extended Werner-like (EWL) state. This is a general state, since it can be reduced to a mixed state or a Bell state depending on the purity $\lambda~\in~[0,1]$ \cite{Bellomo2008}. In general, entanglement becomes limited as purity decreases \cite{Wootters_2001}. The positive partial transpose (PPT) criterion indicates that the Werner state projected on a Bell basis is separable at $\lambda <1/3$ and entangled at $1/3<\lambda \leq1$. The separable decomposition of the EWL state is different from the PS state mentioned above. Here, EWL is represented as a density matrix. Based from Ref.~\cite{Bellomo2008}, we will utilize the expression for EWL as follows
\begin{eqnarray}
    W^{\pm0110}(0)&=&\lambda \ketbra{e^{\pm}_{0110}}+\frac{1-\lambda}{4}I_{4}, \nonumber \\
    W^{\pm0011}(0)&=&\lambda \ketbra{e^{\pm}_{0011}}+\frac{1-\lambda}{4}I_{4},
\end{eqnarray}
where 
\begin{eqnarray}
\label{Bell}
        \ket{e^{\pm}_{0011}}&=& \frac{1}{\sqrt{2}}(\ket{00}\pm\ket{11}), \nonumber \\
        \ket{e^{\pm}_{0110}}&=& \frac{1}{\sqrt{2}}(\ket{01}\pm\ket{10}),
\end{eqnarray}
are Bell states, and $I_{4}$ is a four-dimensional identity matrix. In a matrix representation, the matrix elements of $W^{0110}(0)$ are
\begin{eqnarray}
\label{W0110eq}
    W^{\pm0110}_{11}(0) &=& W^{\pm0110}_{44}= \frac{1-\lambda}{4}, \nonumber \\ 
    W^{\pm0110}_{22}(0) &=& W^{\pm0110}_{33}(0) =\frac{1-\lambda}{4}+\frac{\lambda}{2}, \\
    W^{\pm0110}_{14}(0) &=& 0, \quad W^{\pm0110}_{23}(0)=\frac{\lambda}{2},\nonumber
\end{eqnarray}
where the rest of the elements are zero. On the other hand, $W^{0011}(0)$ has matrix elements
\begin{eqnarray}
\label{W0011eq}
    W^{\pm0011}_{11}(0)&=& W^{\pm0011}_{44}(0)=\frac{1-\lambda}{4}+\frac{\lambda}{2}, \nonumber\\
    W^{\pm0011}_{22}(0)&=& W^{\pm0011}_{33}(0)=\frac{1-\lambda}{4}, \\
    W^{\pm0011}_{14}(0)&=& \frac{\lambda}{2}, \quad W^{\pm0011}_{23}(0)= 0, \nonumber
\end{eqnarray}
and the rest are zero. Consequently, the general form of Werner states follows an ``X" density matrix structure,
\begin{equation}
W(0)=
    \begin{pmatrix}
     W_{11}(0) & 0 & 0 & W_{14}(0)\\
     0 & W_{22}(0) & W_{23}(0) & 0\\
     0 & W_{23}(0) & W_{33}(0) & 0\\
     W_{14}(0) & 0 & 0 & W_{44}(0)
    \end{pmatrix}.
\end{equation}} 
\textcolor{black}{We can see from Eqs. (\ref{W0110eq}) and (\ref{W0011eq}) that EWL states are independent of signs ($\pm$). Then hereon we use $W^{\pm0110}\rightarrow W^{0110}$, and $W^{\pm0011}\rightarrow W^{0011}$ for brevity of notations.}

\section{Results}
\subsection{Mutual qubit environment}\label{Mutual qubit environment}
\begin{figure}[t]
\centering
\includegraphics[width=1.0\linewidth]{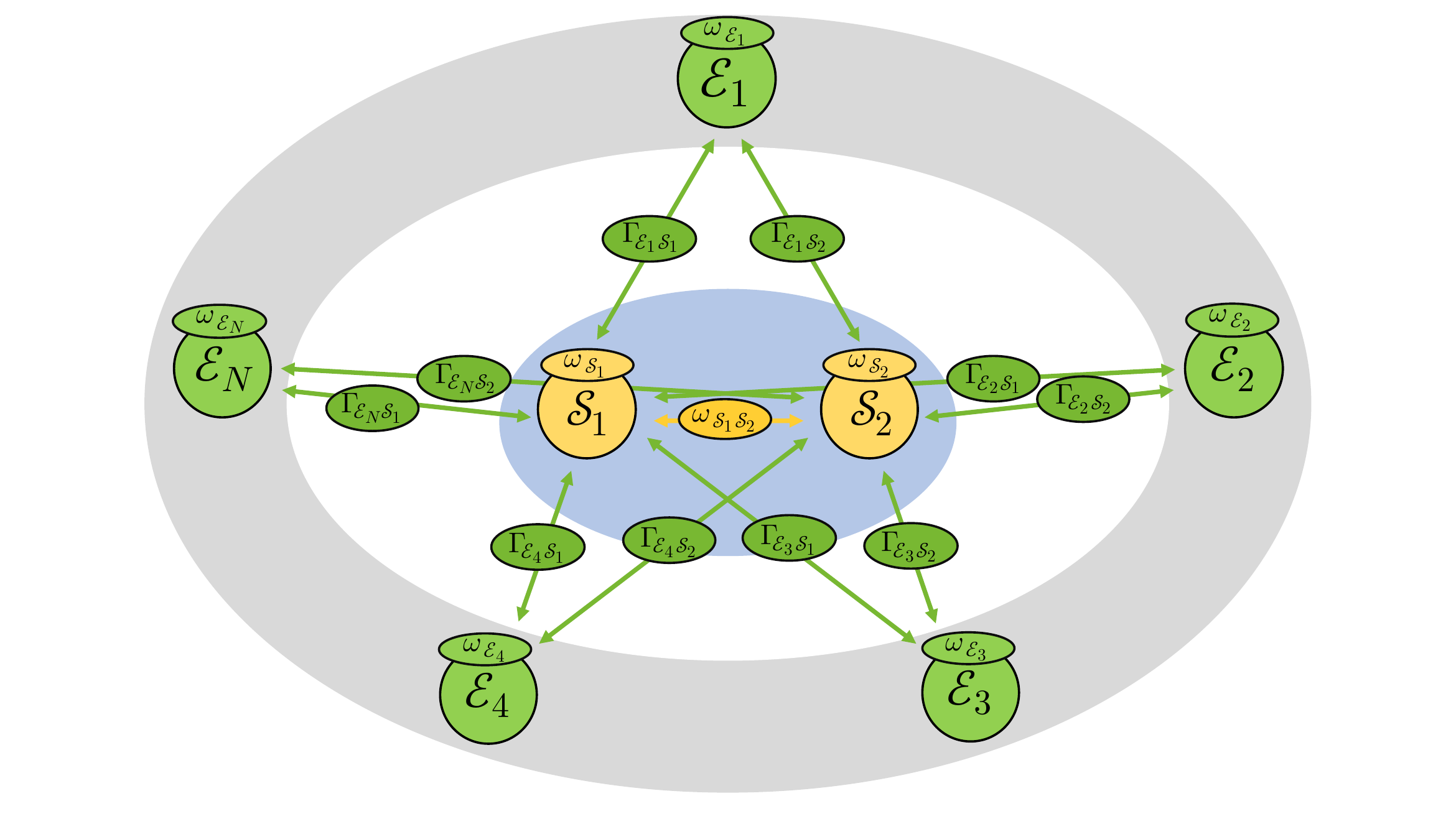}
\caption{\label{Thesis_Model} The system of interest (SOI) comprises of $\mathcal{S}_1$ and $\mathcal{S}_2$ that individually interact with each qubit $\mathcal{E}_i$ in the environment.}
\end{figure}

\textcolor{black}{The two-qubit system, our SOI, consists of two subsystems, $\mathcal{S}_1$ and $\mathcal{S}_2$. Both $\mathcal{S}_1$ and $\mathcal{S}_2$ interact directly with each of the $N$ qubits in the environment $\mathcal{E}$. Each qubit in $\mathcal{E}$ is initially set to be in PS state and unbiased. Figure~\ref{Thesis_Model} illustrates the model.} 

\textcolor{black}{To calculate $\rho_{\mathcal{S}}$, we must first identify the Hilbert spaces involved in the composite system. The Hilbert space is expressed as $\mathcal{H}=(\mathbb{C}_{\mathcal{S}_{1}}^{2}\otimes\mathbb{C}_{\mathcal{S}_{2}}^{2})\bigotimes^{N}_{k=1}\mathbb{C}_{\mathcal{E}_{k}}^{2}$, where $\mathbb{C}_{k}^{2}$ represents the $k^{\text{th}}$ qubit Hilbert space with orthonormal bases $\ket{0}$ and $\ket{1}$. The composite system can be described by the Hamiltonian,
\begin{equation}
    \label{Hterms}\hat{H}=\hat{H}_{\mathcal{S}}+\hat{H}_{\mathcal{E}}+\hat{H}_{\text{int}},
\end{equation}
 where $\hat{H}_{\mathcal{S}}$ is the SOI self-Hamiltonian, $\hat{H}_{\mathcal{E}}$ the self-Hamiltonian of the environment $\mathcal{E}$, and $\hat{H}_{\text{int}}$ the SOI-$\mathcal{E}$ interaction. These are expressed as,
\begin{eqnarray}
\label{HSE}
\hat{H}_{\mathcal{S}}&=& \frac{\hbar}{2}\omS{1}\sigS{1}+\frac{\hbar}{2}\omS{2}\sigS{2},\nonumber\\
\hat{H}_{\E}&=& \frac{\hbar}{2}\sum_{k=1}^{N}\omE{k}\sigE{k}, \nonumber \\
\hat{H}_{\text{int}}&=& \frac{\hbar}{2}\omSS\sigS{1}\sigS{2}  + \frac{\hbar}{2}\sumk \g{1}\sigS{1}\sigE{k} + \frac{\hbar}{2}\sumk \g{2}\sigS{2}\sigE{k}.
\end{eqnarray}
We will limit our parameters $\omS{1}$, $\omS{2}$, $\omE{k}$, $\omSS$, $\g{1}$, and $\g{2}$ to be $\mathbb{R}+$ without loss of generality, while $k=1, 2, 3,\dots , N$ describe the $k$th qubit in $\mathcal{E}$.  Also, $\omSS$ represents the interaction coupling between $\mathcal{S}_{1}$ and $\mathcal{S}_{2}$. Meanwhile, $\g{1}$ and $\g{2}$ refer to the coupling interactions between the $k$th qubit environment and subsystems $\mathcal{S}_{1}$ and $\mathcal{S}_{2}$, respectively.} 

\textcolor{black}{The self-Hamiltonian usually drives the time evolution of the system. The self-Hamiltonian is not considered in the study of Gedik \cite{Gedik2006} since the author focused more on Bell's inequality in coherent regimes. The only thing that changes the coherence of the system is the interaction of the system with the environment. The usual effect of an open quantum system that interacts with an environment is decoherence. However, for the case involving the revival of coherence \cite{Sese2022}, the effect of the self-Hamiltonian on the entanglement of SOI is worth investigating.}

\textcolor{black}{Cucchietti and Zurek \cite{Cucchietti2005} also considered the self-Hamiltonian of SOI.  They used it to determine how the pointer state of SOI emerges based on the competing strength of the self-Hamiltonian and the interaction Hamiltonian with the qubit environment. For this current work, we retain the self-Hamiltonian for the environment $\hat{H}_{\mathcal{E}}$. However, $\hat{H}_{\mathcal{E}}$ has no effect on SOI evolution, since $\hat{H}_{\mathrm{int}}$ commutes with $\hat{H}_{\mathcal{E}}$.} 

\textcolor{black}{For $\hat{H}_{\mathrm{int}}$, there are three terms. The first term corresponds to the interaction of the two qubits of SOI. This term has direct effects on the entanglement time evolution, as will be presented in the succeeding discussions. The second and third terms correspond to the interaction of each individual qubit in SOI with the environment.}   

\textcolor{black}{Let us first consider the initial SOI state as product-separable (PS). This initial state is expressed as,
\begin{equation}
\label{psi start}
    \ket{\Phi(0)}=\ket{\Psi(0)}_{PS}\bigotimes_{k=1}^{N} \ket{\psi_{\mathcal{E}_{k}}(0)}.
\end{equation}
Here, $\ket{\Psi(0)}_{PS}$ is a PS state defined in Eq.~(\ref{PSS_1}). The state vector $\ket{\psi_{\mathcal{E}_{k}}}$ is the initial state of the $k^{\mathrm{th}}$-qubit environment defined as,
\begin{equation}
\label{psi_env}
\ket{\psi_{\mathcal{E}_{k}}(0)}=\alpha_{k}\ket{0}_{k}+\beta_{k}\ket{1}_{k}.
\end{equation}
Here, $\alpha_{k}$ and $\beta_{k}$ are the probability amplitudes of $\ket{0}_{k}$ and $\ket{1}_{k}$, respectively. Considering these bases are eigenstates of $\hat{\sigma}^{z}_{A}$ to an arbitrary Hilbert space $\mathbb{C}^{2}_{A}$, the eigenvalue equations for these bases can be written as,
\begin{equation}
\begin{split}
    \hat{\sigma}^{z}_{A}\ket{0}_{k}=&\ +1\ket{0}_{k},\\
     \hat{\sigma}^{z}_{A}\ket{1}_{k}=&\ -1\ket{1}_{k}.
\end{split}
\end{equation}
We are restricting the entire composite system to be a closed quantum system. Thus, the time evolution of $\ket{\Phi(0)}$ in the Schrödinger picture is $\ket{\Phi(t)}=\mathcal{U}(t)\ket{\Phi(0)}$, where $\mathcal{U}(t)$ is the unitary time evolution operator. Upon substituting Eq.~(\ref{HSE}), the time-evolved state yields
\begin{equation}
\label{CompoState}
    \ket{\Phi(t)}=\sum^{1}_{l,m=0}a_{lm}s_{lm}(t)\ket{lm}\bigotimes_{k=1}^{N}\ket{\psi_{\mathcal{E}_{k,lm}}(t)},
\end{equation}
where the time-evolution of $k^{\text{th}}$-qubit environment $\ket{\psi_{\mathcal{E}_{k,lm}}(t)}$ are as follows
\begin{equation}
\begin{split}
\label{te_env}
    \ket{\psi_{\mathcal{E}_{k}{00\atop 11}}(t)}=&\ \alpha_{k}e^{-it(\omE{k}\pm \gl{1}\pm \gl{2})/2}\ket{0}_{k}+\beta_{k}e^{it(\omE{k}\pm \gl{1}\pm \gl{2})/2}\ket{1}_{k},\\
    \ket{\psi_{\mathcal{E}_{k}{01\atop 10}}(t)}=&\ \alpha_{k} e^{-it(\omE{k}\pm \gl{1}\mp \gl{2})/2}\ket{0}_{k}+\beta_{k}e^{it(\omE{k}\pm \gl{1}\mp \gl{2})/2}\ket{1}_{k}.\\
\end{split}
\end{equation}
Moreover, the time-evolved parameter $s_{lm}(t)$ takes the form
\begin{equation}
\begin{split}
\label{s(t)}
    \s{{00\atop 01}}=&\ \exp\left[-\frac{it}{2}(\omS{1}\pm\omS{2}\pm\omSS)\right],\\
    \s{{10\atop 11}}=&\ \exp\left[\frac{it}{2}(\omS{1}\mp\omS{2}\pm\omSS)\right].
\end{split}
\end{equation}
The upper (lower) indices correspond to the upper (lower) signs.  This notation is used throughout the paper.} 

\textcolor{black}{The density matrix of the entire composite system is the projector of $\ket{\Phi(t)}$,
\begin{equation}
\label{CompoDens}
    \rho_{\mathcal{S}\mathcal{E}}(t)=\sum_{l,m,n,o=0}^{1}a_{lm}a^{*}_{no}s_{lm}(t)s_{no}^{*}(t)\ket{lm}\bra{no}\bigO\ket{\En{_{k,lm}}}\bra{\En{_{k,no}}}.
\end{equation}
Equation (\ref{CompoDens}) can either be a pure or a mixed state. Since the composite system is a closed system, $\rho_{\mathcal{S}\mathcal{E}}(t)$ is always pure. However, this is not necessarily true for SOI since it interacts with $\mathcal{E}$. In fact, we can regard SOI as an open quantum system.}

\textcolor{black}{Since we are only intereseted with the entanglement time evolution of SOI, we must find $\rho_{\mathcal{S}}$ by partial tracing $\rho_{\mathcal{S}\mathcal{E}}$,
\begin{equation}
\label{rho_s(1)}
    \rho_{\mathcal{S}}(t)=\text{Tr}_{\mathcal{E}}[\rho_{\mathcal{S}\mathcal{E}}(t)]=\sum_{P_{1}}\cdots\sum_{P_{N}}{\bra{P_{1}\cdots P_{N}}}\rho_{\mathcal{S}\mathcal{E}}(t)\ket{P_{1}\cdots P_{N}},
\end{equation}
where we define $P_{k}$ as the orthonormal bases $\ket{0}_{k} $ and $ \ket{1}_{k}$ of $k^{\text{th}}$-qubit environment only. Partial tracing $\rho_{\mathcal{SE}}(t)$ will take the form of Eq.~(\ref{rho_S(t)}), which we derived as $a_{lmno}=a_{lm}a_{no}^*$ and $r_{lmno}(t)$ as,
\begin{equation}
    r_{lmno}(t)=s_{lm}(t)s_{no}^{*}(t)\prod^{N}_{k=1}\bigg(\sum^{1}_{P_{k}=0}\bra{P_{k}}\ket{\En{_{k,lm}}}\bra{\En{_{k,no}}}\ket{P_{k}}\bigg) .
\end{equation}
From this, the diagonal elements $r_{llnn}(t)=1$ for all $l$ and $n$. Meanwhile, the decoherence factors (off-diagonal elements) are expressed as
\begin{align}
    \label{r(t)_1}
    r_{\substack{0010\\ 0111}}(t)&= r_{\substack{1000\\ 1101}}^{*}(t)=e^{-it(\omS{1}\pm\omSS)} \prod_{k=1}^{N}\big(|\alpha_{k}|^{2}e^{-it\g{1}}+|\beta_{k}|^{2}e^{it\g{1}}\big),\nonumber \\
    r_{\substack{0001\\ 1011}}(t)&= r_{\substack{0100\\ 1110}}^{*}(t)=e^{-it(\omS{2}\pm\omSS)} \prod_{k=1}^{N}\big(|\alpha_{k}|^{2}e^{-it\g{2}}+|\beta_{k}|^{2}e^{it\g{2}}\big), \\
    r_{\substack{0011\\ 0110}}(t)&= r_{\substack{1100\\ 1001}}^{*}(t)=e^{-it(\omS{1}\pm\omS{2})}\prod_{k=1}^{N}\big(|\alpha_{k}|^{2}e^{-it(\g{1}\pm\g{2})} +|\beta_{k}|^{2}e^{it(\g{1}\pm\g{2})}\big), \nonumber
\end{align}
\noindent These decoherence factors describe the interaction between the subsystems and the environment, which dampens the SOI's interference over time \cite{Zurek1982,Zurek1991}. Similar decoherence factors have been derived in previous studies \cite{Cucchietti2005, Gedik2006, Sese2022, Alporha_2024}.}

\begin{figure}[t]
\begin{subfigure}[t]{0.49\textwidth}
    \includegraphics[width=1.0\linewidth]{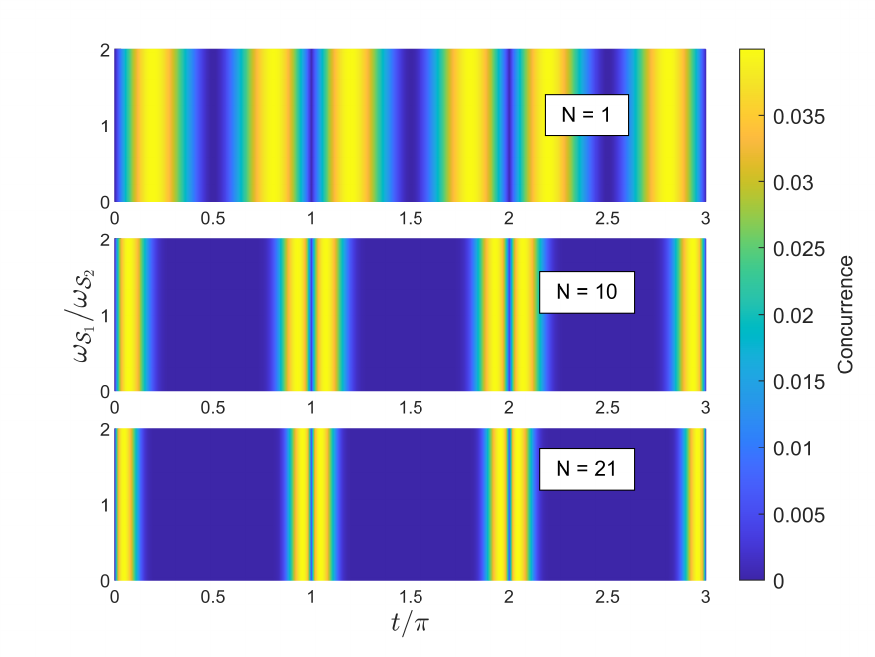}
    \caption{\label{w1w2_Homo_Sep_Post} }
\end{subfigure}
\begin{subfigure}[t]{0.49\textwidth}
    \includegraphics[width=1.0\linewidth]{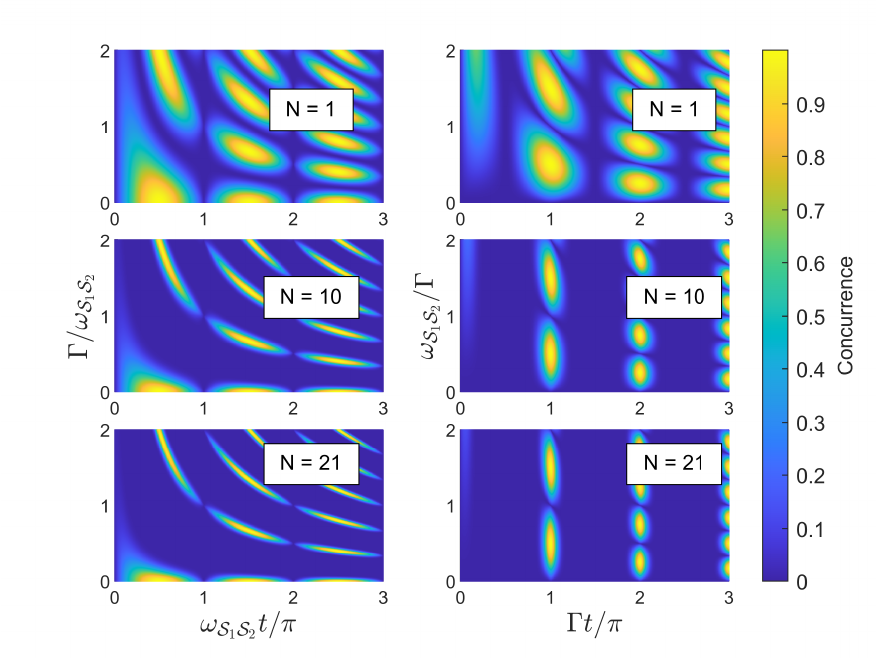}
    \caption{\label{Homo_Sep_Contour_Post} }
\end{subfigure}
\caption{\label{FIG_2} Time evolution of concurrence for an initially prepared PS state. (a) The invariance in the time evolution of concurrence with respect to the internal dynamics $\omS{1}$ of $\mathcal{S}_1$. (b) Time evolution of concurrence with varying interaction strength $\Gamma$ (left column) and varying local interaction $\omSS$ (right column).}
\end{figure}

\begin{figure}[t]
    \includegraphics[width=1.0\linewidth]{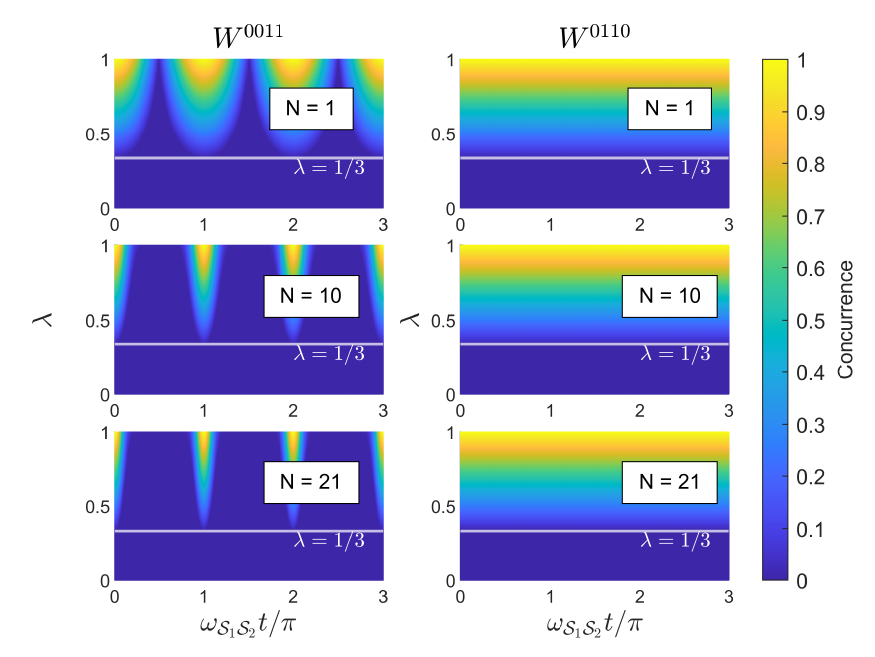}
    \caption{\label{Homo_Werner} Time evolution of concurrence of initially prepared EWL states with varying purity $\lambda$, when $\Gamma/\omSS=0.5$.}
\end{figure}

\begin{figure}[t]
    \includegraphics[width=1.0\linewidth]{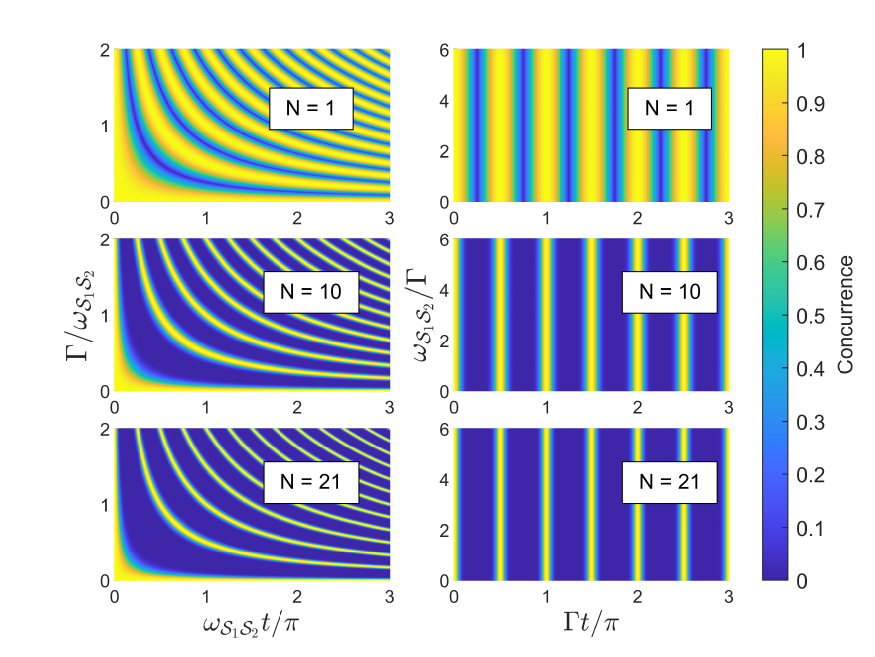}
    \caption{\label{Homo_Bell0011_Post} Time evolution of concurrence in an initially prepared $W^{0011}$ Bell state with varying interaction strength $\Gamma$ (left column) and varying local interaction $\omSS$ (right column).}
\end{figure}

\begin{figure}
    \includegraphics[width=1.0\linewidth]{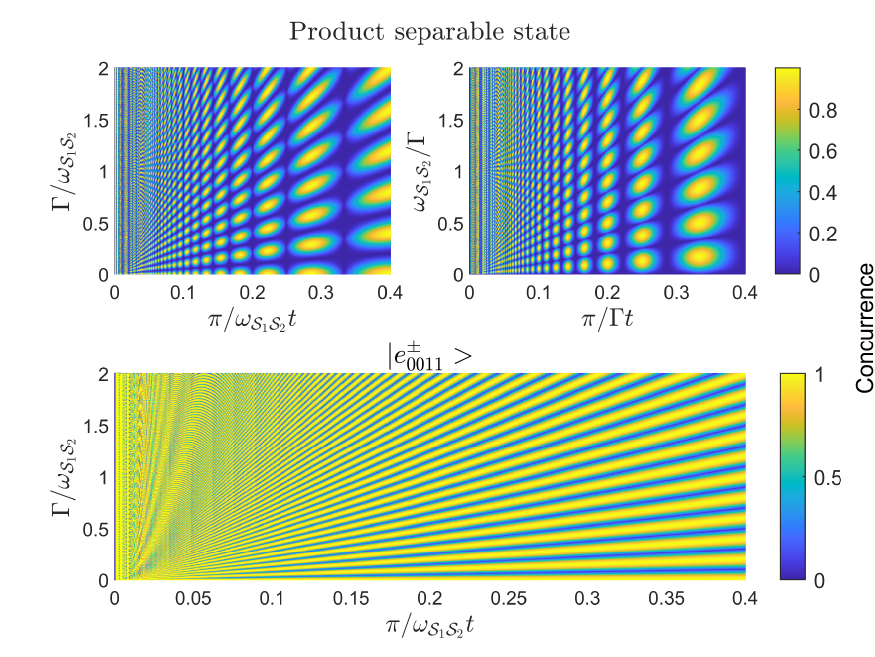}
    \caption{\label{Iso_Post} Linearization of concurrence by setting the horizontal axis to $1/t$ with $N=1$.}
\end{figure}

\textcolor{black}{Considering a homogeneous environment, where we set $\g{1}=\g{2}=\Gamma$ and $|\alpha_{k}|^{2}=|\beta_{k}|^{2}=|\alpha|^{2}=|\beta|^{2}$, we can perform a binomial expansion on Eqs.~(\ref{r(t)_1}) such that it simplifies to
\begin{widetext}
\begin{align}
\label{r(t)_Binomial}
         r_{\substack{0010\\ 0111}}&= r_{\substack{1000\\ 1101}}^{*}=e^{-it(\omS{1} \pm \omSS)}\sum_{k=0}^{N}\begin{pmatrix} N\\k \end{pmatrix}
         |\alpha|^{2k}|\beta|^{2(N-k)}e^{i\Gamma(N-2k)t},\nonumber \\
        r_{\substack{0001\\ 1011}}&=  r_{\substack{0100\\ 1110}}^{*}=e^{-it(\omS{2} \pm \omSS)}\sum_{k=0}^{N}\begin{pmatrix} N\\k \end{pmatrix}|\alpha|^{2k}|\beta|^{2(N-k)}e^{i\Gamma(N-2k)t}, \\
         r_{0011}&= r_{1100}^{*}=e^{-it(\omS{1}+\omS{2})} \sum_{k=0}^{N}\begin{pmatrix} N\\k \end{pmatrix}|\alpha|^{2k}|\beta|^{2(N-k)}e^{2i\Gamma(N-2k)t}, \nonumber \\
         r_{0110}&= r_{1001}^{*}=e^{-it(\omS{1}-\omS{2})}.\nonumber
\end{align}
\end{widetext}
These decoherence factors have the same form as Refs.~\cite{Cucchietti2005,Sese2022}. Sese and Galapon \cite{Sese2022} classified these functions as quasi-periodic functions \cite{Bohr1947,Corduneanu_C}. This means that with these conditions, coherence revival is plausible, which also implies an entanglement revival in time. Such periodic time entanglement revival is consistently observed in Figures~\ref{FIG_2} to \ref{Iso_Post}.} 

\textcolor{black}{In the succeeding discussions $\omSS$, $\Gamma$, and $N$ (number of qubits in the environment) are varied to examine their individual effects. To show the entanglement time evolution, we plot the calculated concurrence using contour maps. The color gradient represents concurrence, where the value of one corresponds to maximally entangled and zero to separable. Throughout this article, the vertical and horizontal axes represent the varied parameter and the rescaled time, respectively. In addition, all axes are rescaled such that the parameter varied is dimensionless.} 

\textcolor{black}{The contour maps of Figure~\ref{FIG_2} represent the time evolution of concurrence for an initially prepared PS state. In Figure~\ref{w1w2_Homo_Sep_Post}, we investigate the effect of varying the parameter $\omS{1}$ that represents the driving frequency of the self-Hamiltonian of subsystem $\mathcal{S}_1$. Here, the vertical axis is rescaled by taking any arbitrary constant value of $\omS{2}$, and the ratio $\Gamma/\omSS{}=1$. It can be seen in this figure that the time evolution of concurrence does not affect the changes in $\omS{1}$. Since SOI is symmetric with respect to permutation between $\mathcal{S}_1$ and $\mathcal{S}_2$, the result is the same when varying $\omS{2}$. This invariance in the time evolution of concurrence is due to the fact that both $\omS{1}$ and $\omS{2}$ are properties of the individual subsystems, and concurrence is a measure of entanglement of SOI. It is important to highlight that entanglement is due to the local interaction between two subsystems. Hence, the self-Hamiltonian of the individual subsystems does not affect the entanglement of SOI. This is also true for the initially prepared EWL states. However, increasing $N$ leads to the attenuation of the time interval that the two qubits in SOI are entangled. This scenario is only true for an initially prepared PS state.}

\textcolor{black}{In the left column of Figure~\ref{Homo_Sep_Contour_Post}, we plot the time evolution of concurrence under varying $\Gamma$. The vertical axis is rescaled by taking any arbitrary constant value of $\omSS$. When $\Gamma \ll \omSS$, entanglement is robust, and does not easily decay. The same is true for rescaling time by taking any arbitrary constant value of $\omSS$. This result suggests that the local interaction $\omSS$ is indeed the driver of entanglement. In contrast, entanglement is easily suppressed by increasing $N$. Every slice of $\Gamma$ creates a different characteristic time evolution for concurrence. We can see that higher values of $\Gamma$ yield shorter period of entanglement revival. This suggests that the interaction with the environment dictates the period of entanglement in SOI. This is verified in the right column of Figure~\ref{Homo_Sep_Contour_Post} by setting an arbitrary constant value of $\Gamma$.} 

\textcolor{black}{Looking at the right column of Figure~\ref{Homo_Sep_Contour_Post}, we vary $\omSS$ (rescaled by an arbitrary constant value of $\Gamma$). Here, we can see that there are values of concurrence~$\sim0$ (gaps) around the neighborhood of $\omSS/\Gamma\sim n$, where $n$ is an integer. This means that it is still possible to have a state that is near to separable even in the presence of local interaction in the SOI as long as these conditions are satisfied.} 

\textcolor{black}{For the second case, where the intial state is EWL, the density matrix of the entire composite system can be written as
\begin{equation}
\label{rho_entire}
    \rho(0)=W(0)\bigotimes_{k=1}^{N} \ketbra{\psi_{\mathcal{E}_{k}}(0)}.
\end{equation}
The time evolution of this general density matrix is obtained by $\rho(t)=\mathcal{U}(t)\rho(0)\mathcal{U}^{\dagger}(t)$. After doing the partial trace, the reduced density matrix of SOI will take the same form in Eq.~(\ref{rho_S(t)}). The coefficients $a_{lmno}$ correspond to the matrix elements of EWL state. Similarly, the same form is obtained for $r_{lmno}(t)$ as describe in Eq.~(\ref{r(t)_1}).}

\textcolor{black}{Applying similar steps of rescaling the vertical and horizontal axes in Figure~\ref{FIG_2}, the effect of varying the purity $\lambda$ is seen in Figure~\ref{Homo_Werner} for an initially prepared $W^{0011}$ and $W^{0110}$ for an arbitrary constant $\omSS$ and $\Gamma$. It can be seen that concurrence is periodic with time for an initially prepared $W^{0011}$, while for $W^{0110}$ it is constant for all time for a particular value of $\lambda$. Here, $\lambda$ sets the subspace of time evolution for an initially prepared EWL state in Hilbert space. From here, we can represent $\lambda=1$ as this type of state is considered maximally entangled.} 

\textcolor{black}{In Figure~\ref{Homo_Bell0011_Post}, a constant curve of concurrence is seen for varying $\Gamma$ and $t$. The time evolution of concurrence of $W^{0110}$ is no longer shown here as it is observed that, just like the entanglement of $W^{0110}$ and $W^{0011}$, it is invariant with respect to $\omSS$. Moreover, unlike $W^{0011}$, which is vulnerable to environment, $W^{0110}$ exhibited immunity to the homogeneous environment: a manifestation of decoherence-free subspace \cite{Lidar1998}. The relationship between $\Gamma$ and the entanglement time evolution of Werner states is similar to that of the PS state, but without the gaps observed. The periodic oscillation of an initially prepared PS state and $W^{0011}$, likewise, having a constant $W^{0110}$ state can be explained by the theory of angular momentum. It is known that the state $W^{0110}$ is an eigenstate of $\hat{\sigma}_{\mathcal{S}_{12}}^{z}=\hat{\sigma}_{\mathcal{S}_{1}}^{z}+\hat{\sigma}_{\mathcal{S}_{2}}^{z}$, and this commutes with the local interaction of SOI considered. This means that this state is a constant of motion for the given Hamiltonian in Eq.~(\ref{HSE}). Unlike $W^{0110}$ state, PS and $W^{0011}$ states are not eigenstates of the $\hat{\sigma}_{\mathcal{S}_{12}}^{z}$. Thus, time periodic oscillation in concurrence for these states is observed.}

\textcolor{black}{
Based from Figures \ref{Homo_Sep_Contour_Post}
and
\ref{Homo_Bell0011_Post},
PS and $W^{0011}$ share similar behaviors. For $\Gamma \gg \omSS$, a much smaller period $\Theta$ for the entanglement revival is observed. We can say that $\Theta$ can be neatly expressed as $\Theta = 2\pi/\Gamma$. Here we can see that $\Gamma$ has an inverse relationship with $\Theta$. Equation (\ref{r(t)_Binomial}) is symmetric with respect to permutation of $\omSS$ and $\Gamma$.  Thus, it has the same relationship for $\omSS$. This motivates us to linearize the contour maps presented in Figure \ref{Iso_Post}.}

\textcolor{black}{In this study, it is worth noting that the time evolution of PS does not attain a maximally entangled state, as can be seen in Figure~\ref{FIG_2}. This means that the subspace for the time evolution of PS state is disjoint with the subspace for the time evolution of EWL state. In other words, the type of model that we consider here cannot generate a maximally entangled state from an initially prepared PS state.} 

\begin{figure}[t]
\begin{subfigure}[t]{0.49\textwidth}
    \includegraphics[width=1.0\linewidth]{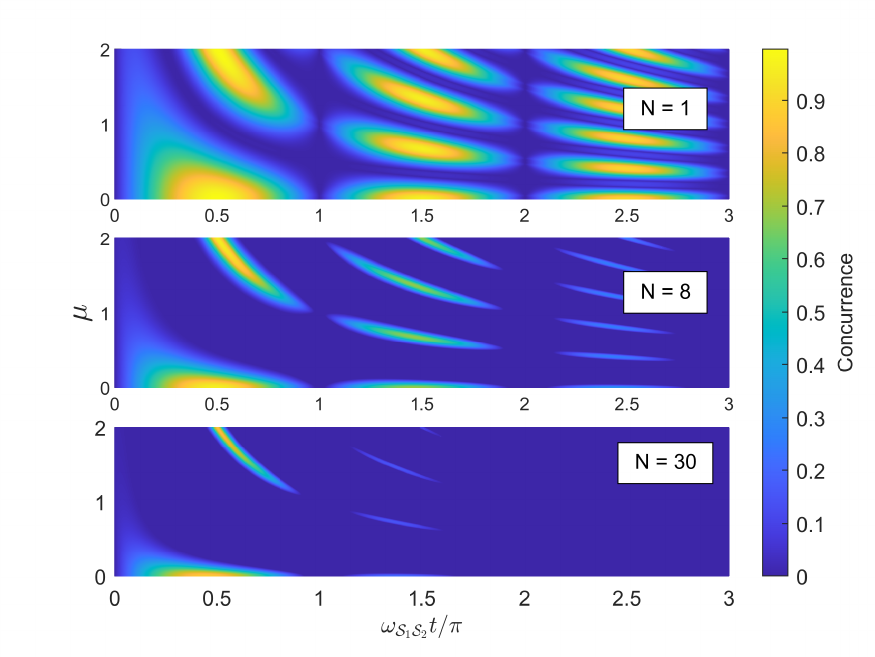}
    \caption{\label{Hetero_Sep_Contour} }
\end{subfigure}
\begin{subfigure}[t]{0.49\textwidth}
    \includegraphics[width=1.0\linewidth]{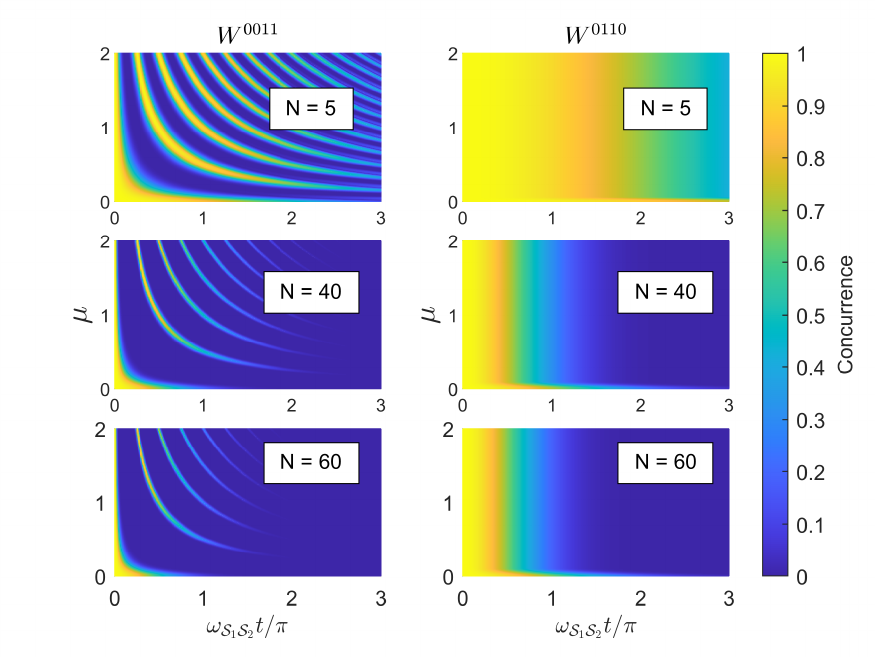}
    \caption{\label{Homo_Bell} }
\end{subfigure}
\caption{\label{FIG_6} Concurrence of SOI interacting with a white noise environment with varying mean $\mu$ where $f=0.1$ is considered. (a) Initially prepared PS state. (b) Initially prepared EWL states.}
\end{figure}

\textcolor{black}{In a chaotic environment such white noise, we use a random number generator (RNG) in MATLAB to randomize $\g{1}$ and $\g{2}$, then setting $\g{i}/\omSS$ as the mean $\mu$ of the distribution with a range $f$ away from $\mu$. Here, $\mu$ can represent the effective interaction strength of the SOI with the qubit environment. In the case where $f < \mu$, we utilize $|\mu - f|$ to prevent any negative interaction strengths in calculations. Since $N$ does not approach the thermodynamic limit, RNG can only generate a sparse set of random numbers within the distribution. Hence, all RNG distributions will just yield similar results.}

\begin{figure}[t]
    \includegraphics[width=1.0\linewidth]{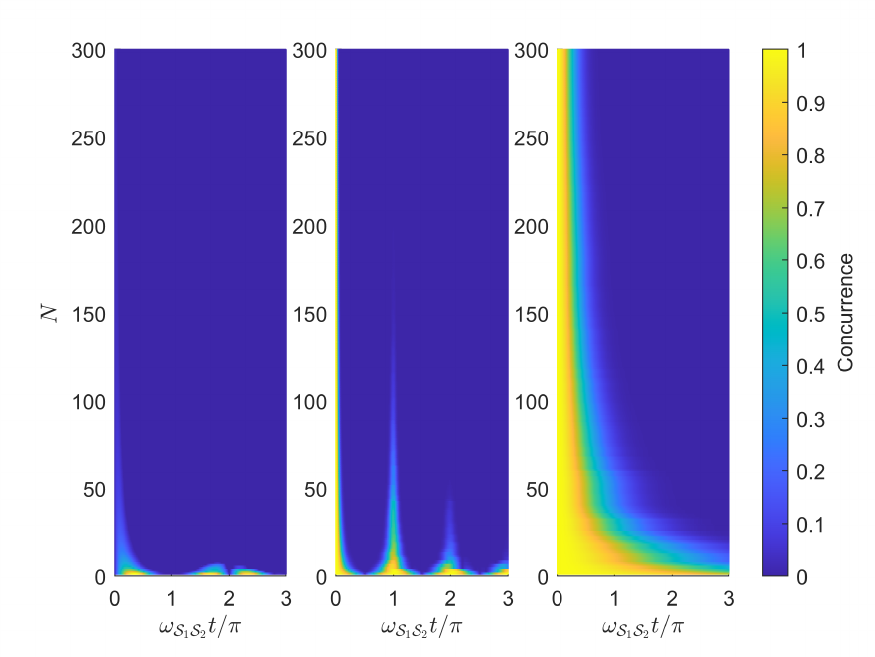}
    \caption{\label{N_Contour} Concurrence with varying $N$ in an initially prepared PS state (left), $W^{0011}$ (middle), and $W^{0110}$ (right) states at $\lambda=1$ with $\mu=0.5$ and $f=0.1$.}
\end{figure}

\textcolor{black}{The results in Figure \ref{FIG_6} share similar characteristics to those previously presented for homogeneous environment. However, these show entanglement decay, which is common in large and chaotic environments. This can be explained by considering a real random number generator with the same distribution. In the range of values for $\g{1(2)}\in[\mu-f,\mu+f]$, the cardinality\footnote{Total number of elements within the set} of irrational numbers is greater than the rational numbers. Then it is more probable that an irrational number is taken for a particular value of $\g{1(2)}$. Having a set of rationally independent interaction couplings induces decay in Eq.~(\ref{r(t)_1}) since there will be no common period for this kind of configuration \cite{Bohr1947,Corduneanu_C}. This also implies an asymptotic entanglement decay for $t\rightarrow\infty$ especially for large $N$. From the study of Cucchietti and Zurek \cite{Cucchietti2005}, it is shown that these decoherence factor in Eq.~(\ref{r(t)_1}) will have a Gaussian decay over time by increasing $N$. For entanglement, this can be seen in Figure \ref{N_Contour}.} 

\subsection{\label{Distinct Environment}Distinct qubit environment}
\textcolor{black}{Up to this point, we only considered a shared environment for both qubit systems. However, it is also possible to consider a Hamiltonian for distinct environments, as illustrated in Figure \ref{Thesis_Model 2}. The Hamiltonian for this model is expressed as,
\begin{align}
\label{HSeparate}
\hat{H}_{\mathcal{S}}&\ = \frac{\hbar}{2}\omS{1}\sigS{1}+\frac{\hbar}{2}\omS{2}\sigS{2},\nonumber &&\\
\hat{H}_{\E}&\ = \frac{\hbar}{2}\sum_{k=1}^{N_{1}}\omES{k}{1}\sigES{k}{1}+\frac{\hbar}{2}\sum_{j=1}^{N_{2}}\omES{j}{2}\sigES{j}{2}, \nonumber &&\\
\hat{H}_{\text{int}}&\ = \frac{\hbar}{2}\omSS\sigS{1}\sigS{2}+\frac{\hbar}{2}\sum_{k=1}^{N_{1}} \gES{k}{1}\sigES{k}{1}\sigS{1} + \frac{\hbar}{2}\sum_{j=1}^{N_{2}} \gES{j}{2}\sigES{j}{2}\sigS{2}. \nonumber
\end{align}
\begin{figure}[t]
    \centering
    \includegraphics[width=1.0\linewidth]{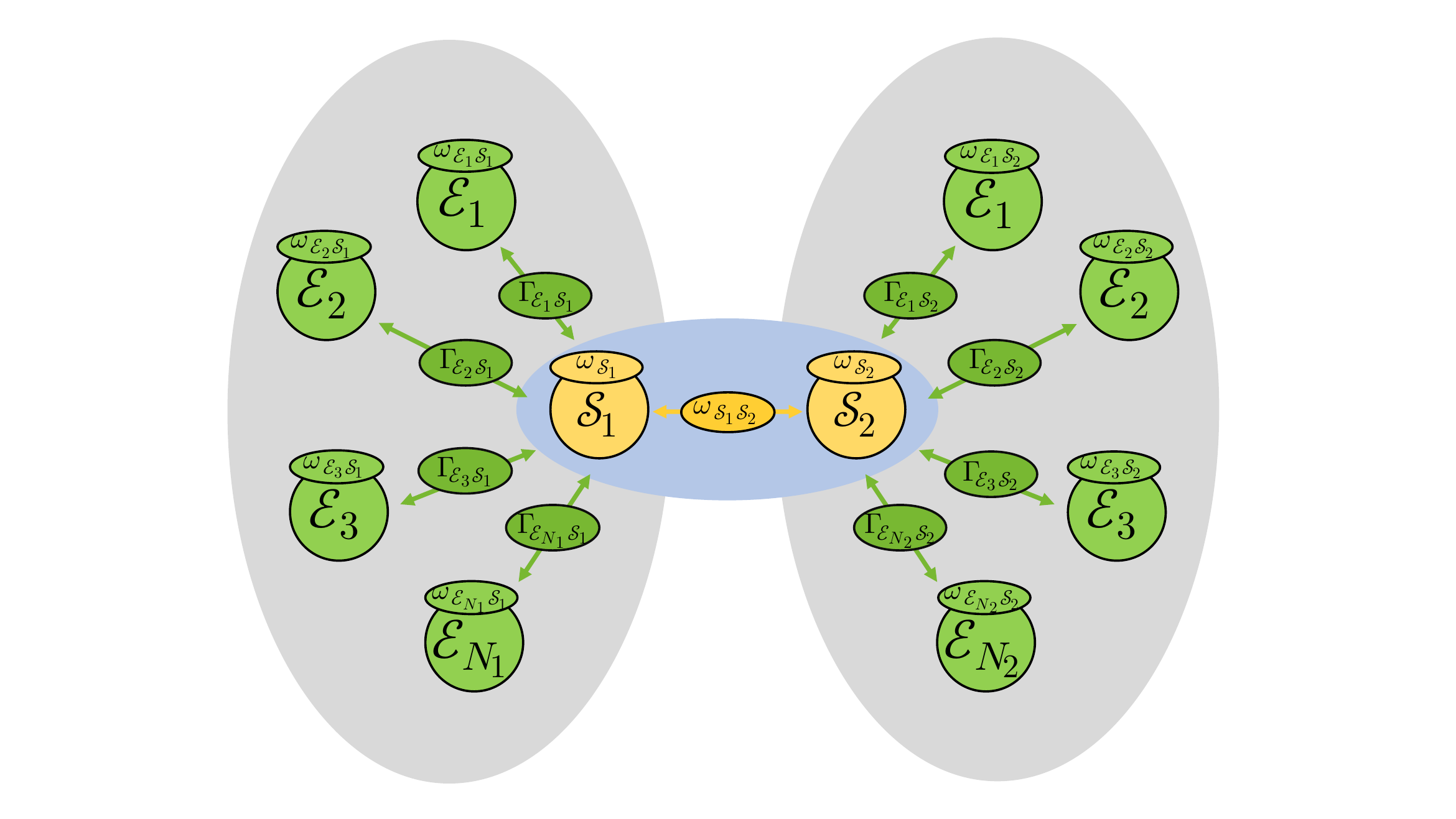}
    \caption{\label{Thesis_Model 2} Two-qubit system interacting in a distinct environment. Here, $\mathcal{S}_1$ and $\mathcal{S}_2$ are interacting independently with their respective environments.} 
\end{figure}
This configuration is also interesting to investigate since, in a realistic scenario, it is possible that the two subsystems have different environments they are interacting with. In this case, the set of environment $\mathcal{E}_{\mathcal{S}_{1}}$ interacting with $\mathcal{S}_{1}$ is different with the set of environment $\mathcal{E}_{\mathcal{S}_{2}}$ interacting with $\mathcal{S}_2$. This allows us to analyze the entanglement dynamics of the SOI when each subsystem is in a distinct environment. Employing similar methodologies in Sec.~\ref{Mutual qubit environment}, the decoherence factors obtained are as follows
\begin{widetext}
    \begin{eqnarray}
    \label{r(t)_Separate}
    r_{\substack{0010\\ 0111}}(t) = r_{\substack{1000\\ 1101}}^{*}(t) &=& e^{-it(\omS{1}\pm\omSS)} \prod_{k=1}^{N_{1}}\big(|\alpha_{k}|^{2}e^{-it\gES{k}{1}}+|\beta_{k}|^{2}e^{it\gES{k}{1}}\big),\nonumber \\
    r_{\substack{0001\\ 1011}}(t) = r_{\substack{0100\\ 1110}}^{*}(t) &=& e^{-it(\omS{2}\pm\omSS)} \prod_{j=1}^{N_{2}}\big(|\alpha_{j}|^{2}e^{-it\gES{j}{2}}+|\beta_{j}|^{2}e^{it\gES{j}{2}}\big),  \\
    r_{\substack{0011\\ 0110}}(t) = r_{\substack{1100\\ 1001}}^{*}(t) &=& e^{-it(\omS{1}\pm\omS{2})}\prod_{k=1}^{N_{1}}\big(|\alpha_{k}|^{2}e^{-it\gES{k}{1}}+|\beta_{k}|^{2}e^{it\gES{k}{1}}\big)\times\nonumber \\
    && \prod_{j=1}^{N_{2}}\big(|\alpha_{j}|^{2}e^{\mp it\gES{j}{2}}+|\beta_{j}|^{2}e^{\pm it\gES{j}{2}}\big)\nonumber,
    \end{eqnarray}
\end{widetext}
where for homogeneous environment we express the environmental interactions as $\gES{k}{1}=\gS{1}$ and $\gES{j}{2}=\gS{2}$.}

\begin{figure}[t]
    \includegraphics[width=1.0\linewidth]{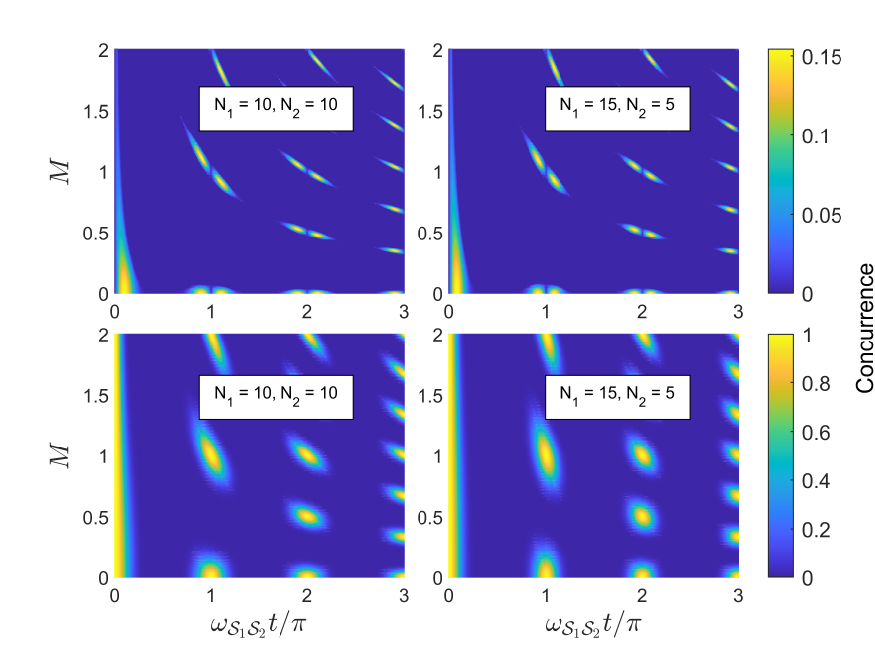}
    \caption{\label{M_N1N2} Concurrence versus $M$ for an initially prepared PS state ($1$st row) and Bell state ($2$nd row) when $\Gamma/\omSS=1$.}
\end{figure}

\textcolor{black}{In this section, we consider two different environmental scenarios: $\gS{1}= M \gS{2}$, where $M$ is a positive real number, to examine if it is possible to reproduce the results in a shared environment from a distinct environment; and lastly, when one is homogeneous while the other is a white noise environment.}

\subsubsection{\label{Case 1} Case 1: $\gS{1}=M\gS{2}$}
\begin{figure}[t]
\begin{subfigure}[t]{0.49\textwidth}
    \includegraphics[width=1.0\linewidth]{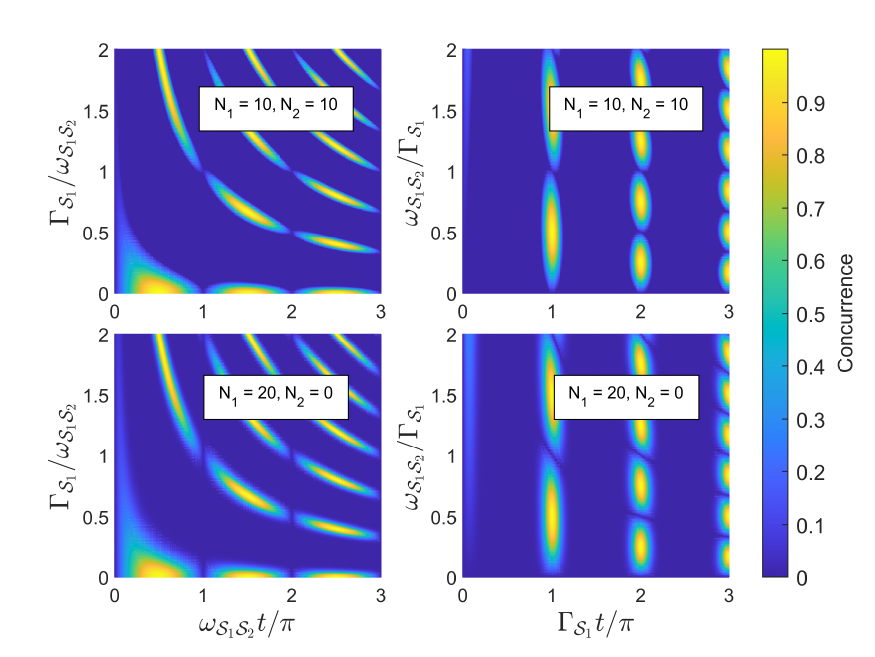}
    \caption{\label{Homo_Sep_N1N2} }
\end{subfigure}
\begin{subfigure}[t]{0.49\textwidth}
    \includegraphics[width=1.0\linewidth]{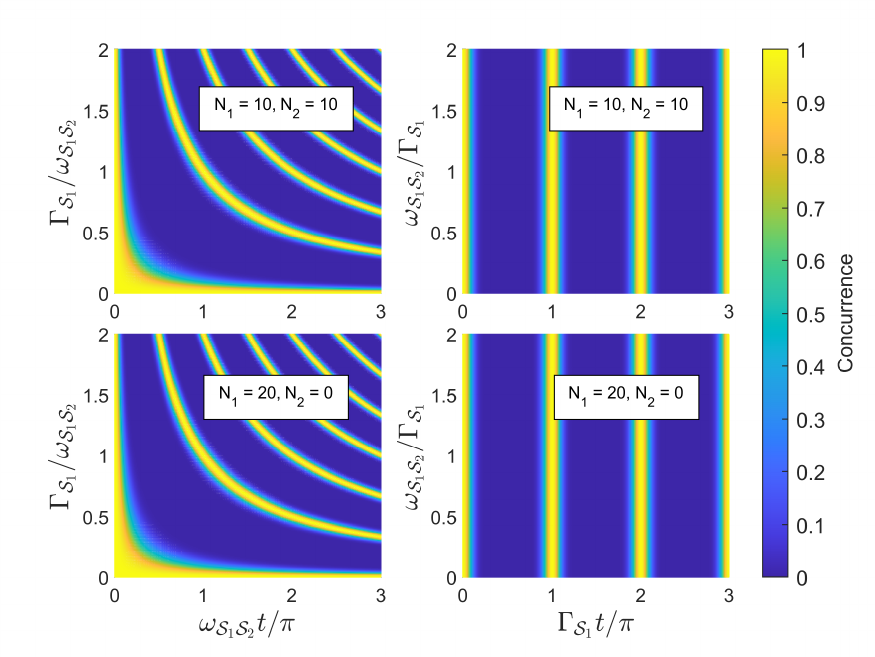}
    \caption{\label{Homo_Bell_N1N2} }
\end{subfigure}
\caption{\label{FIG_10} Concurrence time evolution when $M=1$. (a) Initially prepared PS state. (b) Initially prepared $W^{0011}$ state.}
\end{figure}

\textcolor{black}{By separating the environment, we observe in Figures \ref{M_N1N2} to \ref{Homo_Bell_N1N2} a different entanglement time evolution compared to the mutual environment. For instance, when $N=10$ in a mutual environment, we have to set $M=1$ and $N_{1}=N_{2}=10$ in a distinct environment to equalize the interaction strengths and the number of environmental interactions. However, this approach results in a larger product iteration in Eq. (\ref{r(t)_Separate}) compared to Eq. (\ref{r(t)_1}), leading to a different entanglement time evolution.}

\textcolor{black}{Although the symmetry of the entanglement time evolution with respect to $N$ is trivial, having the same total number of environments does not necessarily result in identical entanglement time evolution for an initially prepared PS state. That is, the entanglement time evolution depends on the distribution of qubit environments interacting with each subsystem (see Figure \ref{Homo_Sep_N1N2}). In contrast, in Figure \ref{Homo_Bell_N1N2}, initially prepared $W^{0011}$ entanglement time evolution is identical in a distinct environment, even with different qubit environment distribution. This occurs because the concurrence of Bell states only depends on the modulus of the decoherence factors, rendering the relative spin orientation trivial \cite{Gedik2006}.}
\begin{figure}[t]
    \includegraphics[width=1.0\linewidth]{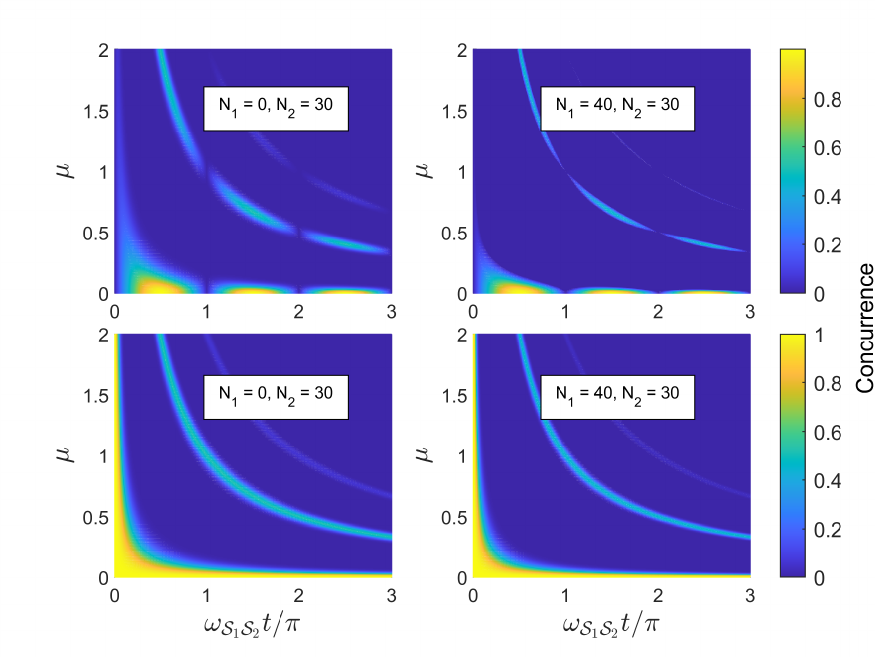}
    \caption{\label{Hetero_Sep_Bell_N1N2} Concurrence time evolution with PS state ($1$st row) and Bell state ($2$nd row) as their initial states for $f=0.1\mu $ and $M=1$.}
\end{figure}
 
\subsubsection{\label{Case 2} Case 2: $\gS{1}$ and $\gES{j}{2}$}
\textcolor{black}{Another possible scenario involves a distinct environment where one of the two environments is homogeneous while the other is white noise. As shown in Figure \ref{Hetero_Sep_Bell_N1N2}, the dissipation time decreases as the range $f$ increases, since increasing $f$ provides a larger cardinality in the given set of distribution. Furthermore, introducing a homogeneous environment ($N_1$) to SOI with a white noise environment ($N_2$) further attenuates the concurrence, with only a minimal decrease in decay time. This result is expected, as attenuation predominantly occurs in homogeneous environments, while entanglement decay occurs in white noise environments.}

\section{\label{Conclusion}Conclusion}
\textcolor{black}{We studied the entanglement time evolution of two distinct initially prepared states, namely PS and EWL states, in both mutual and distinct qubit environment. In a pure homogeneous environment, both states demonstrate complete entanglement revival, where an increase in $N$ leads to an attenuation of concurrence. Additionally, the entanglement time evolution of the PS state depends on $\Gamma$ and $\omSS$, whereas only on $\Gamma$ for EWL states.  The PS state exhibits gaps and never reaches a maximally entangled state, whereas the states $W^{0110}$ and $W^{0011}$ maintain and periodically reach maximum entanglement for $\lambda=1$, respectively. Hence, the PS and EWL states possess unique properties, making their entanglement time evolution subspaces disjoint from each other. This result indicates that it is impossible to generate a Bell state from an initially prepared PS state for the model we considered. We also found that when $\Gamma$ or $\omSS$ influences entanglement, their relationship with $1/t$ exhibits linearity under a constant value of concurrence. Conversely, placing the states in a white noise environment results in entanglement dissipation, where the Gaussian dissipation time decreases as $N$ increases. Interestingly, we found that the dissipation time depends on the width of the interaction strength random distribution $f$ rather than its value.}

\textcolor{black}{In contrast, placing SOI in a distinct qubit environment reveals that the distribution of the environment influences the entanglement time evolution of the PS state, while the EWL state is independent of the environment's distribution. Furthermore, combining homogeneous and white noise environment results in a mixture of homogeneous and white noise characteristics into the SOI entanglement time evolution. These findings provide valuable insights into identifying a state based on its entanglement time evolution in a homogeneous or white noise environment, even without knowledge of the system's initial preparation. In the case of an infinite qubit bath, one can employ the similar methodologies done by Cucchietti and Zurek \cite{Cucchietti2005}. However, this condition is already enough to encapsulate the behavior of the system when SOI is interacting with an infinite qubit bath.}

\textcolor{black}{This work provides a simplified model of how the qubit environment influences the system's entanglement. One possible extension is to consider a mutually interacting environment, an arbitrary bipartite system, a multipartite system, a comparison of coherence and entanglement, and a simple application to quantum circuits.}

\bibliography{References}

\begin{thebibliography}{34}%
\makeatletter
\providecommand \@ifxundefined [1]{%
 \@ifx{#1\undefined}
}%
\providecommand \@ifnum [1]{%
 \ifnum #1\expandafter \@firstoftwo
 \else \expandafter \@secondoftwo
 \fi
}%
\providecommand \@ifx [1]{%
 \ifx #1\expandafter \@firstoftwo
 \else \expandafter \@secondoftwo
 \fi
}%
\providecommand \natexlab [1]{#1}%
\providecommand \enquote  [1]{``#1''}%
\providecommand \bibnamefont  [1]{#1}%
\providecommand \bibfnamefont [1]{#1}%
\providecommand \citenamefont [1]{#1}%
\providecommand \href@noop [0]{\@secondoftwo}%
\providecommand \href [0]{\begingroup \@sanitize@url \@href}%
\providecommand \@href[1]{\@@startlink{#1}\@@href}%
\providecommand \@@href[1]{\endgroup#1\@@endlink}%
\providecommand \@sanitize@url [0]{\catcode `\\12\catcode `\$12\catcode `\&12\catcode `\#12\catcode `\^12\catcode `\_12\catcode `\%12\relax}%
\providecommand \@@startlink[1]{}%
\providecommand \@@endlink[0]{}%
\providecommand \url  [0]{\begingroup\@sanitize@url \@url }%
\providecommand \@url [1]{\endgroup\@href {#1}{\urlprefix }}%
\providecommand \urlprefix  [0]{URL }%
\providecommand \Eprint [0]{\href }%
\providecommand \doibase [0]{https://doi.org/}%
\providecommand \selectlanguage [0]{\@gobble}%
\providecommand \bibinfo  [0]{\@secondoftwo}%
\providecommand \bibfield  [0]{\@secondoftwo}%
\providecommand \translation [1]{[#1]}%
\providecommand \BibitemOpen [0]{}%
\providecommand \bibitemStop [0]{}%
\providecommand \bibitemNoStop [0]{.\EOS\space}%
\providecommand \EOS [0]{\spacefactor3000\relax}%
\providecommand \BibitemShut  [1]{\csname bibitem#1\endcsname}%
\let\auto@bib@innerbib\@empty
\bibitem [{\citenamefont {Schrödinger}(1935)}]{Schrodinger_1935}%
  \BibitemOpen
  \bibfield  {author} {\bibinfo {author} {\bibfnamefont {E.}~\bibnamefont {Schrödinger}},\ }\bibfield  {title} {\bibinfo {title} {Discussion of probability relations between separated systems},\ }\href {https://doi.org/10.1017/S0305004100013554} {\bibfield  {journal} {\bibinfo  {journal} {Math. Proc. Cambridge Philos. Soc.}\ }\textbf {\bibinfo {volume} {31}},\ \bibinfo {pages} {555–563} (\bibinfo {year} {1935})}\BibitemShut {NoStop}%
\bibitem [{\citenamefont {Ekert}(1991)}]{Ekert1991}%
  \BibitemOpen
  \bibfield  {author} {\bibinfo {author} {\bibfnamefont {A.~K.}\ \bibnamefont {Ekert}},\ }\bibfield  {title} {\bibinfo {title} {Quantum cryptography based on {B}ell's theorem},\ }\href {https://doi.org/10.1103/PhysRevLett.67.661} {\bibfield  {journal} {\bibinfo  {journal} {Phys. Rev. Lett.}\ }\textbf {\bibinfo {volume} {67}},\ \bibinfo {pages} {661} (\bibinfo {year} {1991})}\BibitemShut {NoStop}%
\bibitem [{\citenamefont {Naik}\ \emph {et~al.}(2000)\citenamefont {Naik}, \citenamefont {Peterson}, \citenamefont {White}, \citenamefont {Berglund},\ and\ \citenamefont {Kwiat}}]{Naik2000}%
  \BibitemOpen
  \bibfield  {author} {\bibinfo {author} {\bibfnamefont {D.~S.}\ \bibnamefont {Naik}}, \bibinfo {author} {\bibfnamefont {C.~G.}\ \bibnamefont {Peterson}}, \bibinfo {author} {\bibfnamefont {A.~G.}\ \bibnamefont {White}}, \bibinfo {author} {\bibfnamefont {A.~J.}\ \bibnamefont {Berglund}},\ and\ \bibinfo {author} {\bibfnamefont {P.~G.}\ \bibnamefont {Kwiat}},\ }\bibfield  {title} {\bibinfo {title} {Entangled state quantum cryptography: Eavesdropping on the {E}kert protocol},\ }\href {https://doi.org/10.1103/PhysRevLett.84.4733} {\bibfield  {journal} {\bibinfo  {journal} {Phys. Rev. Lett.}\ }\textbf {\bibinfo {volume} {84}},\ \bibinfo {pages} {4733} (\bibinfo {year} {2000})}\BibitemShut {NoStop}%
\bibitem [{\citenamefont {Tittel}\ \emph {et~al.}(2000)\citenamefont {Tittel}, \citenamefont {Brendel}, \citenamefont {Zbinden},\ and\ \citenamefont {Gisin}}]{Tittel2000}%
  \BibitemOpen
  \bibfield  {author} {\bibinfo {author} {\bibfnamefont {W.}~\bibnamefont {Tittel}}, \bibinfo {author} {\bibfnamefont {J.}~\bibnamefont {Brendel}}, \bibinfo {author} {\bibfnamefont {H.}~\bibnamefont {Zbinden}},\ and\ \bibinfo {author} {\bibfnamefont {N.}~\bibnamefont {Gisin}},\ }\bibfield  {title} {\bibinfo {title} {Quantum cryptography using entangled photons in energy-time bell states},\ }\href {https://doi.org/10.1103/PhysRevLett.84.4737} {\bibfield  {journal} {\bibinfo  {journal} {Phys. Rev. Lett.}\ }\textbf {\bibinfo {volume} {84}},\ \bibinfo {pages} {4737} (\bibinfo {year} {2000})}\BibitemShut {NoStop}%
\bibitem [{\citenamefont {Mattle}\ \emph {et~al.}(1996)\citenamefont {Mattle}, \citenamefont {Weinfurter}, \citenamefont {Kwiat},\ and\ \citenamefont {Zeilinger}}]{Mattle1996}%
  \BibitemOpen
  \bibfield  {author} {\bibinfo {author} {\bibfnamefont {K.}~\bibnamefont {Mattle}}, \bibinfo {author} {\bibfnamefont {H.}~\bibnamefont {Weinfurter}}, \bibinfo {author} {\bibfnamefont {P.~G.}\ \bibnamefont {Kwiat}},\ and\ \bibinfo {author} {\bibfnamefont {A.}~\bibnamefont {Zeilinger}},\ }\bibfield  {title} {\bibinfo {title} {Dense coding in experimental quantum communication},\ }\href {https://doi.org/10.1103/PhysRevLett.76.4656} {\bibfield  {journal} {\bibinfo  {journal} {Phys. Rev. Lett.}\ }\textbf {\bibinfo {volume} {76}},\ \bibinfo {pages} {4656} (\bibinfo {year} {1996})}\BibitemShut {NoStop}%
\bibitem [{\citenamefont {Fang}\ \emph {et~al.}(2000)\citenamefont {Fang}, \citenamefont {Zhu}, \citenamefont {Feng}, \citenamefont {Mao},\ and\ \citenamefont {Du}}]{Fang2000}%
  \BibitemOpen
  \bibfield  {author} {\bibinfo {author} {\bibfnamefont {X.}~\bibnamefont {Fang}}, \bibinfo {author} {\bibfnamefont {X.}~\bibnamefont {Zhu}}, \bibinfo {author} {\bibfnamefont {M.}~\bibnamefont {Feng}}, \bibinfo {author} {\bibfnamefont {X.}~\bibnamefont {Mao}},\ and\ \bibinfo {author} {\bibfnamefont {F.}~\bibnamefont {Du}},\ }\bibfield  {title} {\bibinfo {title} {Experimental implementation of dense coding using nuclear magnetic resonance},\ }\href {https://doi.org/10.1103/PhysRevA.61.022307} {\bibfield  {journal} {\bibinfo  {journal} {Phys. Rev. A}\ }\textbf {\bibinfo {volume} {61}},\ \bibinfo {pages} {022307} (\bibinfo {year} {2000})}\BibitemShut {NoStop}%
\bibitem [{\citenamefont {Mizuno}\ \emph {et~al.}(2005)\citenamefont {Mizuno}, \citenamefont {Wakui}, \citenamefont {Furusawa},\ and\ \citenamefont {Sasaki}}]{Mizuno2005}%
  \BibitemOpen
  \bibfield  {author} {\bibinfo {author} {\bibfnamefont {J.}~\bibnamefont {Mizuno}}, \bibinfo {author} {\bibfnamefont {K.}~\bibnamefont {Wakui}}, \bibinfo {author} {\bibfnamefont {A.}~\bibnamefont {Furusawa}},\ and\ \bibinfo {author} {\bibfnamefont {M.}~\bibnamefont {Sasaki}},\ }\bibfield  {title} {\bibinfo {title} {Experimental demonstration of entanglement-assisted coding using a two-mode squeezed vacuum state},\ }\href {https://doi.org/10.1103/PhysRevA.71.012304} {\bibfield  {journal} {\bibinfo  {journal} {Phys. Rev. A}\ }\textbf {\bibinfo {volume} {71}},\ \bibinfo {pages} {012304} (\bibinfo {year} {2005})}\BibitemShut {NoStop}%
\bibitem [{\citenamefont {Jing}\ \emph {et~al.}(2003)\citenamefont {Jing}, \citenamefont {Zhang}, \citenamefont {Yan}, \citenamefont {Zhao}, \citenamefont {Xie},\ and\ \citenamefont {Peng}}]{Jing2003}%
  \BibitemOpen
  \bibfield  {author} {\bibinfo {author} {\bibfnamefont {J.}~\bibnamefont {Jing}}, \bibinfo {author} {\bibfnamefont {J.}~\bibnamefont {Zhang}}, \bibinfo {author} {\bibfnamefont {Y.}~\bibnamefont {Yan}}, \bibinfo {author} {\bibfnamefont {F.}~\bibnamefont {Zhao}}, \bibinfo {author} {\bibfnamefont {C.}~\bibnamefont {Xie}},\ and\ \bibinfo {author} {\bibfnamefont {K.}~\bibnamefont {Peng}},\ }\bibfield  {title} {\bibinfo {title} {Experimental demonstration of tripartite entanglement and controlled dense coding for continuous variables},\ }\href {https://doi.org/10.1103/PhysRevLett.90.167903} {\bibfield  {journal} {\bibinfo  {journal} {Phys. Rev. Lett.}\ }\textbf {\bibinfo {volume} {90}},\ \bibinfo {pages} {167903} (\bibinfo {year} {2003})}\BibitemShut {NoStop}%
\bibitem [{\citenamefont {Bennett}\ \emph {et~al.}(1993)\citenamefont {Bennett}, \citenamefont {Brassard}, \citenamefont {Cr\'epeau}, \citenamefont {Jozsa}, \citenamefont {Peres},\ and\ \citenamefont {Wootters}}]{Bennett1993}%
  \BibitemOpen
  \bibfield  {author} {\bibinfo {author} {\bibfnamefont {C.~H.}\ \bibnamefont {Bennett}}, \bibinfo {author} {\bibfnamefont {G.}~\bibnamefont {Brassard}}, \bibinfo {author} {\bibfnamefont {C.}~\bibnamefont {Cr\'epeau}}, \bibinfo {author} {\bibfnamefont {R.}~\bibnamefont {Jozsa}}, \bibinfo {author} {\bibfnamefont {A.}~\bibnamefont {Peres}},\ and\ \bibinfo {author} {\bibfnamefont {W.~K.}\ \bibnamefont {Wootters}},\ }\bibfield  {title} {\bibinfo {title} {Teleporting an unknown quantum state via dual classical and einstein-podolsky-rosen channels},\ }\href {https://doi.org/10.1103/PhysRevLett.70.1895} {\bibfield  {journal} {\bibinfo  {journal} {Phys. Rev. Lett.}\ }\textbf {\bibinfo {volume} {70}},\ \bibinfo {pages} {1895} (\bibinfo {year} {1993})}\BibitemShut {NoStop}%
\bibitem [{\citenamefont {Bouwmeester}\ \emph {et~al.}(1997)\citenamefont {Bouwmeester}, \citenamefont {Pan}, \citenamefont {Mattle}, \citenamefont {Eibl}, \citenamefont {Weinfurter},\ and\ \citenamefont {Zeilinger}}]{Bouwmeester1997}%
  \BibitemOpen
  \bibfield  {author} {\bibinfo {author} {\bibfnamefont {D.}~\bibnamefont {Bouwmeester}}, \bibinfo {author} {\bibfnamefont {J.-W.}\ \bibnamefont {Pan}}, \bibinfo {author} {\bibfnamefont {K.}~\bibnamefont {Mattle}}, \bibinfo {author} {\bibfnamefont {M.}~\bibnamefont {Eibl}}, \bibinfo {author} {\bibfnamefont {H.}~\bibnamefont {Weinfurter}},\ and\ \bibinfo {author} {\bibfnamefont {A.}~\bibnamefont {Zeilinger}},\ }\bibfield  {title} {\bibinfo {title} {Experimental quantum teleportation},\ }\href {https://doi.org/10.1038/37539} {\bibfield  {journal} {\bibinfo  {journal} {Nature}\ }\textbf {\bibinfo {volume} {390}},\ \bibinfo {pages} {575–579} (\bibinfo {year} {1997})}\BibitemShut {NoStop}%
\bibitem [{\citenamefont {Sherson}\ \emph {et~al.}(2006)\citenamefont {Sherson}, \citenamefont {Krauter}, \citenamefont {Olsson}, \citenamefont {Julsgaard}, \citenamefont {Hammerer}, \citenamefont {Cirac},\ and\ \citenamefont {Polzik}}]{Sherson2006}%
  \BibitemOpen
  \bibfield  {author} {\bibinfo {author} {\bibfnamefont {J.~F.}\ \bibnamefont {Sherson}}, \bibinfo {author} {\bibfnamefont {H.}~\bibnamefont {Krauter}}, \bibinfo {author} {\bibfnamefont {R.~K.}\ \bibnamefont {Olsson}}, \bibinfo {author} {\bibfnamefont {B.}~\bibnamefont {Julsgaard}}, \bibinfo {author} {\bibfnamefont {K.}~\bibnamefont {Hammerer}}, \bibinfo {author} {\bibfnamefont {I.}~\bibnamefont {Cirac}},\ and\ \bibinfo {author} {\bibfnamefont {E.~S.}\ \bibnamefont {Polzik}},\ }\bibfield  {title} {\bibinfo {title} {Quantum teleportation between light and matter},\ }\href {https://doi.org/10.1038/nature05136} {\bibfield  {journal} {\bibinfo  {journal} {Nature}\ }\textbf {\bibinfo {volume} {443}},\ \bibinfo {pages} {557–560} (\bibinfo {year} {2006})}\BibitemShut {NoStop}%
\bibitem [{\citenamefont {Jozsa}\ and\ \citenamefont {Linden}(2003)}]{Linden2002}%
  \BibitemOpen
  \bibfield  {author} {\bibinfo {author} {\bibfnamefont {R.}~\bibnamefont {Jozsa}}\ and\ \bibinfo {author} {\bibfnamefont {N.}~\bibnamefont {Linden}},\ }\bibfield  {title} {\bibinfo {title} {On the role of entanglement in quantum-computational speed-up},\ }\href {https://doi.org/10.1098/rspa.2002.1097} {\bibfield  {journal} {\bibinfo  {journal} {Proc. R. Soc. A: Math. Phys. Eng. Sci.}\ }\textbf {\bibinfo {volume} {459}},\ \bibinfo {pages} {2011} (\bibinfo {year} {2003})}\BibitemShut {NoStop}%
\bibitem [{\citenamefont {Or\'us}\ and\ \citenamefont {Latorre}(2004)}]{Orus2004}%
  \BibitemOpen
  \bibfield  {author} {\bibinfo {author} {\bibfnamefont {R.}~\bibnamefont {Or\'us}}\ and\ \bibinfo {author} {\bibfnamefont {J.~I.}\ \bibnamefont {Latorre}},\ }\bibfield  {title} {\bibinfo {title} {Universality of entanglement and quantum-computation complexity},\ }\href {https://doi.org/10.1103/PhysRevA.69.052308} {\bibfield  {journal} {\bibinfo  {journal} {Phys. Rev. A}\ }\textbf {\bibinfo {volume} {69}},\ \bibinfo {pages} {052308} (\bibinfo {year} {2004})}\BibitemShut {NoStop}%
\bibitem [{\citenamefont {Horodecki}\ \emph {et~al.}(2009)\citenamefont {Horodecki}, \citenamefont {Horodecki}, \citenamefont {Horodecki},\ and\ \citenamefont {Horodecki}}]{Horodecki_2009}%
  \BibitemOpen
  \bibfield  {author} {\bibinfo {author} {\bibfnamefont {R.}~\bibnamefont {Horodecki}}, \bibinfo {author} {\bibfnamefont {P.}~\bibnamefont {Horodecki}}, \bibinfo {author} {\bibfnamefont {M.}~\bibnamefont {Horodecki}},\ and\ \bibinfo {author} {\bibfnamefont {K.}~\bibnamefont {Horodecki}},\ }\bibfield  {title} {\bibinfo {title} {Quantum entanglement},\ }\href {https://doi.org/10.1103/RevModPhys.81.865} {\bibfield  {journal} {\bibinfo  {journal} {Rev. Mod. Phys.}\ }\textbf {\bibinfo {volume} {81}},\ \bibinfo {pages} {865} (\bibinfo {year} {2009})}\BibitemShut {NoStop}%
\bibitem [{\citenamefont {Gottesman}(1998)}]{gottesman1998}%
  \BibitemOpen
  \bibfield  {author} {\bibinfo {author} {\bibfnamefont {D.}~\bibnamefont {Gottesman}},\ }\href {https://arxiv.org/abs/quant-ph/9807006} {\bibinfo {title} {The {H}eisenberg representation of quantum computers}} (\bibinfo {year} {1998}),\ \Eprint {https://arxiv.org/abs/quant-ph/9807006} {arXiv:quant-ph/9807006 [quant-ph]} \BibitemShut {NoStop}%
\bibitem [{\citenamefont {Nielsen}\ and\ \citenamefont {Chuang}(2010)}]{Nielsen_et_al_2010}%
  \BibitemOpen
  \bibfield  {author} {\bibinfo {author} {\bibfnamefont {M.~A.}\ \bibnamefont {Nielsen}}\ and\ \bibinfo {author} {\bibfnamefont {I.~L.}\ \bibnamefont {Chuang}},\ }\href@noop {} {\emph {\bibinfo {title} {Quantum Computation and Quantum Information}}}\ (\bibinfo  {publisher} {Cambridge University Press},\ \bibinfo {address} {New York, United States of America},\ \bibinfo {year} {2010})\BibitemShut {NoStop}%
\bibitem [{\citenamefont {\ifmmode~\dot{Z}\else \.{Z}\fi{}yczkowski}\ \emph {et~al.}(2001)\citenamefont {\ifmmode~\dot{Z}\else \.{Z}\fi{}yczkowski}, \citenamefont {Horodecki}, \citenamefont {Horodecki},\ and\ \citenamefont {Horodecki}}]{Zyczkowski_2001}%
  \BibitemOpen
  \bibfield  {author} {\bibinfo {author} {\bibfnamefont {K.}~\bibnamefont {\ifmmode~\dot{Z}\else \.{Z}\fi{}yczkowski}}, \bibinfo {author} {\bibfnamefont {P.}~\bibnamefont {Horodecki}}, \bibinfo {author} {\bibfnamefont {M.}~\bibnamefont {Horodecki}},\ and\ \bibinfo {author} {\bibfnamefont {R.}~\bibnamefont {Horodecki}},\ }\bibfield  {title} {\bibinfo {title} {Dynamics of quantum entanglement},\ }\href {https://doi.org/10.1103/PhysRevA.65.012101} {\bibfield  {journal} {\bibinfo  {journal} {Phys. Rev. A}\ }\textbf {\bibinfo {volume} {65}},\ \bibinfo {pages} {012101} (\bibinfo {year} {2001})}\BibitemShut {NoStop}%
\bibitem [{\citenamefont {Tahira}\ \emph {et~al.}(2008)\citenamefont {Tahira}, \citenamefont {Ikram}, \citenamefont {Azim},\ and\ \citenamefont {Zubairy}}]{Tahira2008}%
  \BibitemOpen
  \bibfield  {author} {\bibinfo {author} {\bibfnamefont {R.}~\bibnamefont {Tahira}}, \bibinfo {author} {\bibfnamefont {M.}~\bibnamefont {Ikram}}, \bibinfo {author} {\bibfnamefont {T.}~\bibnamefont {Azim}},\ and\ \bibinfo {author} {\bibfnamefont {M.~S.}\ \bibnamefont {Zubairy}},\ }\bibfield  {title} {\bibinfo {title} {Entanglement dynamics of a pure bipartite system in dissipative environments},\ }\href {https://doi.org/10.1088/0953-4075/41/20/205501} {\bibfield  {journal} {\bibinfo  {journal} {J. Phys. B-At. Mol. Opt.}\ }\textbf {\bibinfo {volume} {41}},\ \bibinfo {pages} {205501} (\bibinfo {year} {2008})}\BibitemShut {NoStop}%
\bibitem [{\citenamefont {Shi}\ \emph {et~al.}(2016)\citenamefont {Shi}, \citenamefont {Wang},\ and\ \citenamefont {Ye}}]{Shi2016}%
  \BibitemOpen
  \bibfield  {author} {\bibinfo {author} {\bibfnamefont {J.-d.}\ \bibnamefont {Shi}}, \bibinfo {author} {\bibfnamefont {D.}~\bibnamefont {Wang}},\ and\ \bibinfo {author} {\bibfnamefont {L.}~\bibnamefont {Ye}},\ }\bibfield  {title} {\bibinfo {title} {Entanglement revive and information flow within the decoherent environment},\ }\href {https://doi.org/10.1038/srep30710} {\bibfield  {journal} {\bibinfo  {journal} {Sci. Rep.}\ }\textbf {\bibinfo {volume} {6}},\ \bibinfo {pages} {30710} (\bibinfo {year} {2016})}\BibitemShut {NoStop}%
\bibitem [{\citenamefont {Yu}\ and\ \citenamefont {Eberly}(2004)}]{Yu2004}%
  \BibitemOpen
  \bibfield  {author} {\bibinfo {author} {\bibfnamefont {T.}~\bibnamefont {Yu}}\ and\ \bibinfo {author} {\bibfnamefont {J.~H.}\ \bibnamefont {Eberly}},\ }\bibfield  {title} {\bibinfo {title} {Finite-time disentanglement via spontaneous emission},\ }\href {https://doi.org/10.1103/PhysRevLett.93.140404} {\bibfield  {journal} {\bibinfo  {journal} {Phys. Rev. Lett.}\ }\textbf {\bibinfo {volume} {93}},\ \bibinfo {pages} {140404} (\bibinfo {year} {2004})}\BibitemShut {NoStop}%
\bibitem [{\citenamefont {Flores}\ and\ \citenamefont {Galapon}(2015)}]{Flores2015}%
  \BibitemOpen
  \bibfield  {author} {\bibinfo {author} {\bibfnamefont {M.}~\bibnamefont {Flores}}\ and\ \bibinfo {author} {\bibfnamefont {E.}~\bibnamefont {Galapon}},\ }\bibfield  {title} {\bibinfo {title} {Two qubit entanglement preservation through the addition of qubits},\ }\href {https://doi.org/https://doi.org/10.1016/j.aop.2014.11.011} {\bibfield  {journal} {\bibinfo  {journal} {Ann. Phys.}\ }\textbf {\bibinfo {volume} {354}},\ \bibinfo {pages} {21} (\bibinfo {year} {2015})}\BibitemShut {NoStop}%
\bibitem [{\citenamefont {Cucchietti}\ \emph {et~al.}(2005)\citenamefont {Cucchietti}, \citenamefont {Paz},\ and\ \citenamefont {Zurek}}]{Cucchietti2005}%
  \BibitemOpen
  \bibfield  {author} {\bibinfo {author} {\bibfnamefont {F.~M.}\ \bibnamefont {Cucchietti}}, \bibinfo {author} {\bibfnamefont {J.~P.}\ \bibnamefont {Paz}},\ and\ \bibinfo {author} {\bibfnamefont {W.~H.}\ \bibnamefont {Zurek}},\ }\bibfield  {title} {\bibinfo {title} {Decoherence from spin environments},\ }\href {https://doi.org/10.1103/PhysRevA.72.052113} {\bibfield  {journal} {\bibinfo  {journal} {Phys. Rev. A}\ }\textbf {\bibinfo {volume} {72}},\ \bibinfo {pages} {052113} (\bibinfo {year} {2005})}\BibitemShut {NoStop}%
\bibitem [{\citenamefont {Sese}\ and\ \citenamefont {Galapon}(2022)}]{Sese2022}%
  \BibitemOpen
  \bibfield  {author} {\bibinfo {author} {\bibfnamefont {L.~J.~F.}\ \bibnamefont {Sese}}\ and\ \bibinfo {author} {\bibfnamefont {E.~A.}\ \bibnamefont {Galapon}},\ }\bibfield  {title} {\bibinfo {title} {Exact recurrence of a qubit immersed in a homogeneous n-qubit environment},\ }in\ \href {https://proceedings.spp-online.org/article/view/SPP-2022-1D-02} {\emph {\bibinfo {booktitle} {Proceedings of the Samahang Pisika ng Pilipinas}}},\ Vol.~\bibinfo {volume} {40}\ (\bibinfo {year} {2022})\ pp.\ \bibinfo {pages} {SPP--2022--1D--02}\BibitemShut {NoStop}%
\bibitem [{\citenamefont {Gedik}(2006)}]{Gedik2006}%
  \BibitemOpen
  \bibfield  {author} {\bibinfo {author} {\bibfnamefont {Z.}~\bibnamefont {Gedik}},\ }\bibfield  {title} {\bibinfo {title} {Spin bath decoherence of quantum entanglement},\ }\href {https://doi.org/https://doi.org/10.1016/j.ssc.2006.02.004} {\bibfield  {journal} {\bibinfo  {journal} {Solid State Commun.}\ }\textbf {\bibinfo {volume} {138}},\ \bibinfo {pages} {82} (\bibinfo {year} {2006})}\BibitemShut {NoStop}%
\bibitem [{\citenamefont {Alporha}\ \emph {et~al.}(2024)\citenamefont {Alporha}, \citenamefont {Sese},\ and\ \citenamefont {Gammag}}]{Alporha_2024}%
  \BibitemOpen
  \bibfield  {author} {\bibinfo {author} {\bibfnamefont {R.}~\bibnamefont {Alporha}}, \bibinfo {author} {\bibfnamefont {L.~J.}\ \bibnamefont {Sese}},\ and\ \bibinfo {author} {\bibfnamefont {R.}~\bibnamefont {Gammag}},\ }\bibfield  {title} {\bibinfo {title} {Dynamical concurrence of initially separable bipartite system immersed in n-qubit environment},\ }\href {https://doi.org/10.1088/1742-6596/2793/1/012015} {\bibfield  {journal} {\bibinfo  {journal} {J. Phys. Conf. Ser.}\ }\textbf {\bibinfo {volume} {2793}},\ \bibinfo {pages} {012015} (\bibinfo {year} {2024})}\BibitemShut {NoStop}%
\bibitem [{\citenamefont {Zurek}(1991)}]{Zurek1991}%
  \BibitemOpen
  \bibfield  {author} {\bibinfo {author} {\bibfnamefont {W.~H.}\ \bibnamefont {Zurek}},\ }\bibfield  {title} {\bibinfo {title} {{Decoherence and the Transition from Quantum to Classical}},\ }\href {https://doi.org/10.1063/1.881293} {\bibfield  {journal} {\bibinfo  {journal} {Physics Today}\ }\textbf {\bibinfo {volume} {44}},\ \bibinfo {pages} {36} (\bibinfo {year} {1991})}\BibitemShut {NoStop}%
\bibitem [{\citenamefont {Hill}\ and\ \citenamefont {Wootters}(1997)}]{Hill_1997}%
  \BibitemOpen
  \bibfield  {author} {\bibinfo {author} {\bibfnamefont {S.~A.}\ \bibnamefont {Hill}}\ and\ \bibinfo {author} {\bibfnamefont {W.~K.}\ \bibnamefont {Wootters}},\ }\bibfield  {title} {\bibinfo {title} {Entanglement of a pair of quantum bits},\ }\href {https://doi.org/https://doi.org/10.1103/PhysRevLett.78.5022} {\bibfield  {journal} {\bibinfo  {journal} {Phys. Rev. Lett.}\ }\textbf {\bibinfo {volume} {78}},\ \bibinfo {pages} {5022} (\bibinfo {year} {1997})}\BibitemShut {NoStop}%
\bibitem [{\citenamefont {Wootters}(1998)}]{Wootters_1998}%
  \BibitemOpen
  \bibfield  {author} {\bibinfo {author} {\bibfnamefont {W.~K.}\ \bibnamefont {Wootters}},\ }\bibfield  {title} {\bibinfo {title} {Entanglement of formation of an arbitrary state of two qubits},\ }\href {https://doi.org/10.1103/PhysRevLett.80.2245} {\bibfield  {journal} {\bibinfo  {journal} {Phys. Rev. Lett.}\ }\textbf {\bibinfo {volume} {80}},\ \bibinfo {pages} {2245} (\bibinfo {year} {1998})}\BibitemShut {NoStop}%
\bibitem [{\citenamefont {Bellomo}\ \emph {et~al.}(2008)\citenamefont {Bellomo}, \citenamefont {Lo~Franco},\ and\ \citenamefont {Compagno}}]{Bellomo2008}%
  \BibitemOpen
  \bibfield  {author} {\bibinfo {author} {\bibfnamefont {B.}~\bibnamefont {Bellomo}}, \bibinfo {author} {\bibfnamefont {R.}~\bibnamefont {Lo~Franco}},\ and\ \bibinfo {author} {\bibfnamefont {G.}~\bibnamefont {Compagno}},\ }\bibfield  {title} {\bibinfo {title} {Entanglement dynamics of two independent qubits in environments with and without memory},\ }\href {http://dx.doi.org/10.1103/PhysRevA.77.032342} {\bibfield  {journal} {\bibinfo  {journal} {Phys. Rev. A}\ }\textbf {\bibinfo {volume} {77}} (\bibinfo {year} {2008})}\BibitemShut {NoStop}%
\bibitem [{\citenamefont {Wootters}(2001)}]{Wootters_2001}%
  \BibitemOpen
  \bibfield  {author} {\bibinfo {author} {\bibfnamefont {W.~K.}\ \bibnamefont {Wootters}},\ }\bibfield  {title} {\bibinfo {title} {Entanglement of formation and concurrence},\ }\href {https://doi.org/10.26421/QIC1.1-3} {\bibfield  {journal} {\bibinfo  {journal} {Quantum Info. Comput.}\ }\textbf {\bibinfo {volume} {1}},\ \bibinfo {pages} {27–44} (\bibinfo {year} {2001})}\BibitemShut {NoStop}%
\bibitem [{\citenamefont {Zurek}(1982)}]{Zurek1982}%
  \BibitemOpen
  \bibfield  {author} {\bibinfo {author} {\bibfnamefont {W.~H.}\ \bibnamefont {Zurek}},\ }\bibfield  {title} {\bibinfo {title} {Environment-induced superselection rules},\ }\href {https://doi.org/10.1103/PhysRevD.26.1862} {\bibfield  {journal} {\bibinfo  {journal} {Phys. Rev. D}\ }\textbf {\bibinfo {volume} {26}},\ \bibinfo {pages} {1862} (\bibinfo {year} {1982})}\BibitemShut {NoStop}%
\bibitem [{\citenamefont {Bohr}\ and\ \citenamefont {Cohn}(1951)}]{Bohr1947}%
  \BibitemOpen
  \bibfield  {author} {\bibinfo {author} {\bibfnamefont {H.}~\bibnamefont {Bohr}}\ and\ \bibinfo {author} {\bibfnamefont {H.}~\bibnamefont {Cohn}},\ }\href {https://books.google.com.ph/books?id=LkTvAAAAMAAJ} {\emph {\bibinfo {title} {Almost Periodic Functions}}}\ (\bibinfo  {publisher} {Chelsea Publishing Company},\ \bibinfo {year} {1951})\BibitemShut {NoStop}%
\bibitem [{\citenamefont {Corduneanu}(1968)}]{Corduneanu_C}%
  \BibitemOpen
  \bibfield  {author} {\bibinfo {author} {\bibfnamefont {C.}~\bibnamefont {Corduneanu}},\ }\href {https://books.google.com.ph/books?id=-T3tvwEACAAJ} {\emph {\bibinfo {title} {Almost Periodic Functions}}}\ (\bibinfo  {publisher} {Interscience Publishers},\ \bibinfo {year} {1968})\BibitemShut {NoStop}%
\bibitem [{\citenamefont {Lidar}\ \emph {et~al.}(1998)\citenamefont {Lidar}, \citenamefont {Chuang},\ and\ \citenamefont {Whaley}}]{Lidar1998}%
  \BibitemOpen
  \bibfield  {author} {\bibinfo {author} {\bibfnamefont {D.~A.}\ \bibnamefont {Lidar}}, \bibinfo {author} {\bibfnamefont {I.~L.}\ \bibnamefont {Chuang}},\ and\ \bibinfo {author} {\bibfnamefont {K.~B.}\ \bibnamefont {Whaley}},\ }\bibfield  {title} {\bibinfo {title} {Decoherence-free subspaces for quantum computation},\ }\href {https://doi.org/10.1103/PhysRevLett.81.2594} {\bibfield  {journal} {\bibinfo  {journal} {Phys. Rev. Lett.}\ }\textbf {\bibinfo {volume} {81}},\ \bibinfo {pages} {2594} (\bibinfo {year} {1998})}\BibitemShut {NoStop}%
\end{thebibliography}%

\end{document}